\documentclass[aps,prl,showpacs,10pt,twocolumn,superscriptaddress]{revtex4-2}
\usepackage[unicode,breaklinks]{hyperref}
\usepackage{physics,amsmath,amssymb}
\usepackage{dcolumn}
\usepackage{scalerel}
\usepackage{bm}
\usepackage{graphicx}
\usepackage{tikz}
\usetikzlibrary{shapes}
\usepackage{soul}

\graphicspath{ {./images/} }

\begin{document}

\title{Scalable Spin Squeezing in Power-Law Interacting XXZ Models with Disorder}

\author{Samuel E. Begg}
\email{samuel.begg@utdallas.edu}
\affiliation{Department of Physics, The University of Texas at Dallas, Richardson, Texas 75080, USA}
\author{Bishal K. Ghosh}
\affiliation{Department of Physics, The University of Texas at Dallas, Richardson, Texas 75080, USA}
\affiliation{Institute of Physics, University of Graz, Universitätsplatz 5, 8010 Graz, Austria}
\author{Chong Zu}
\author{Chuanwei Zhang}
\affiliation{Department of Physics, Washington University, St. Louis, Missouri 63130, USA}
\affiliation{Center for Quantum Leaps, Washington University, St. Louis, Missouri 63130, USA}
\author{Michael Kolodrubetz}
\affiliation{Department of Physics, The University of Texas at Dallas, Richardson, Texas 75080, USA}

\date{\today}

\begin{abstract}
While spin squeezing has been traditionally considered in all-to-all interacting models, recent works have shown that it can also occur in systems with power-law interactions, enabling direct tests in Rydberg atoms, trapped ions, ultracold atoms, and nitrogen-vacancy (NV) centers in diamond. For the latter, Ref. \cite{Wu2025} demonstrated that spin squeezing is heavily affected by positional disorder, reducing any capacity for a practical squeezing advantage, which requires scalability with the system size. In this Letter we explore the robustness of spin squeezing in two-dimensional lattices with a fraction of unoccupied lattice sites. Using semiclassical modeling, we demonstrate the existence of scalable squeezing in power-law interacting XXZ models up to a disorder threshold, above which squeezing is not scalable. We produce a phase diagram for scalable squeezing, and explain its absence in the aforementioned NV experiment. Our work illustrates the maximum disorder allowed for realizing scalable spin squeezing in a host of quantum simulators, highlights a regime with substantial tolerance to disorder, and identifies controlled defect creation as a promising route for scalable squeezing in solid-state systems.
\end{abstract} 
\maketitle

   Quantum simulators now provide opportunities for engineering collective quantum phenomena with applications to quantum metrology \cite{Degen2017,Pezze2018}. 
  A paradigmatic example  is spin squeezing \cite{Wineland1992,Kitagawa1993,Wineland1994}, whereby the quantum projection noise is reduced (squeezed) in a particular direction, which can be exploited to perform precision measurements.
     While spin squeezing has been traditionally considered in all-to-all interacting models \cite{Kitagawa1993,Ma2011}, recent works have shown that scalable spin squeezing -- where squeezing scales with system size -- is possible in systems with power-law interactions \cite{FossFeig2016,Perlin2020}.
   In particular, dynamics thermalizing to the easy-plane ferromagnetic phase  display scalable spin squeezing 
       \cite{Block_2024}, establishing a link between dynamical scaling of quantum information and equilibrium order. Subsequent work has generalized this to the case of quasi-long range ordered phases \cite{Roscilde2024}. The squeezing dynamics can also be 
       understood from the perspective of 
    a partial decoupling \cite{Roscilde2022,Roscilde2023} of a collective large-spin (Dicke) manifold, in which states exhibit  one-axis twisting (OAT) dynamics  up to finite-temperature corrections \cite{Block_2024}, the simplest form of spin squeezing. For sufficiently long-range interactions, the collective dynamics are protected by a spectral gap \cite{Perlin2020}, an intuition that  has motivated generalizations to  two-mode squeezing \cite{Bilitewski2023a,Duha2024,Duha2025} and two-axis counter-twisting \cite{Koyluoglu2025}.
      
 The ubiquity of power-law interacting systems in quantum simulators has led to recent demonstrations of spin squeezing in Rydberg atoms \cite{Bornet2023a, Eckner2023, Hines2023}, trapped ions \cite{Franke2023a}, neutral atoms with contact \cite{Lee2025} and long-range interactions \cite{Douglas2025}, and nitrogen vacancy (NV) centers  in diamond \cite{Wu2025}. Hamiltonians capable of spin squeezing have also been realized in polar molecules \cite{Miller2024} and cavity-mediated (all-to-all) interactions \cite{Luo2025}. 
    For the NV experiment in Ref.~\cite{Wu2025}, squeezing was shown to be significantly affected by positional disorder of the spins, and was not scalable, despite experimental methods for removing strongly interacting spins  which reduce collective behavior. Similarly, spin squeezing in three-dimensional optical lattices has been shown to be impacted by a finite hole fraction \cite{Lee2025}. 
    A systematic understanding of the robustness of scalable spin squeezing to disorder is currently lacking.
 
 In this Letter we characterize the impact of disorder on spin squeezing by considering regular two-dimensional (2D) lattices with a fraction $p$ of randomly positioned unoccupied lattice sites (vacancies). Using semiclassical modeling, we demonstrate the existence of scalable squeezing in power-law interacting XXZ models up to a disorder threshold, above which the squeezing is not scalable. The transition is found to align with a change in the presence/absence of order at late times, as anticipated from the U(1) symmetry \cite{Block_2024}. Our work illustrates a minimal disorder requirement for realizing scalable spin squeezing in a host of quantum simulators, and explains why Ref. \cite{Wu2025} did not observe scalable squeezing.
We highlight favorable regions of the phase diagram to target in future experiments, and conclude with a discussion of new experimental directions with the potential  to mitigate the effect of the positional disorder.

\begin{figure}[t]
    \centering  \includegraphics[width=0.493\linewidth]{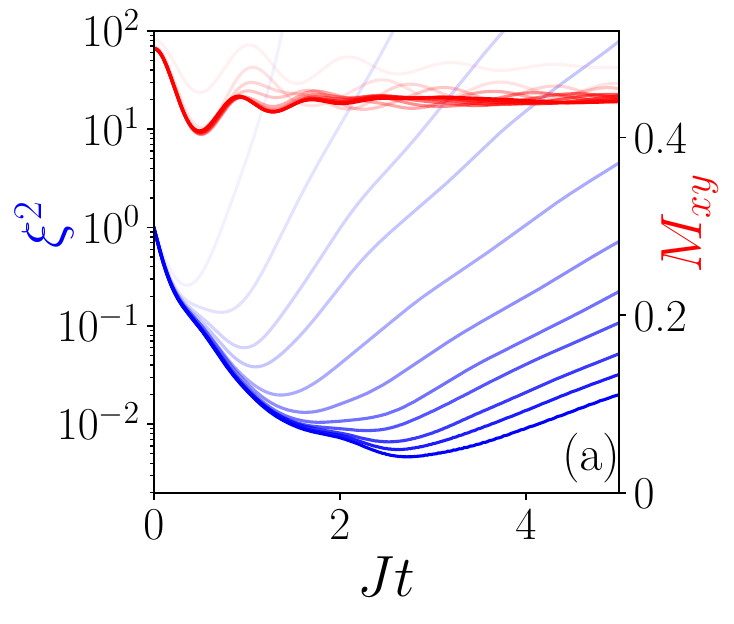}    \includegraphics[width=0.493\linewidth]{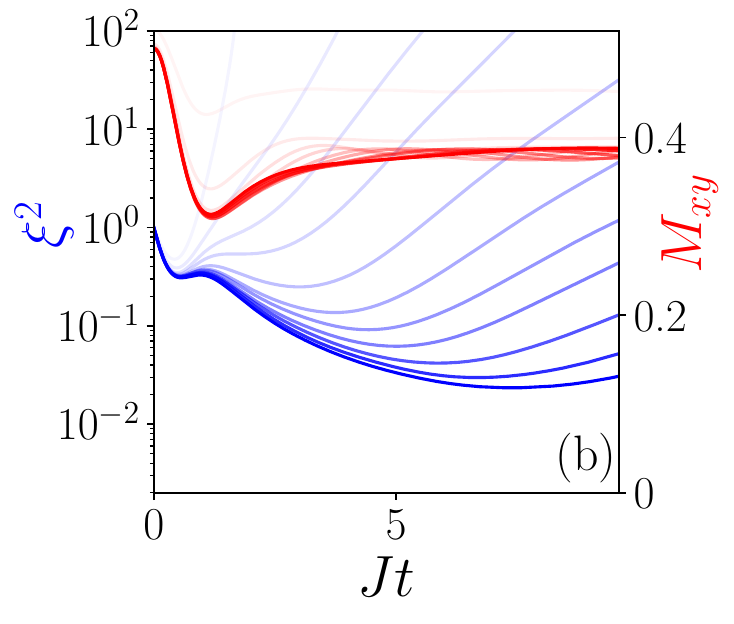}
\includegraphics[width=0.493\linewidth]{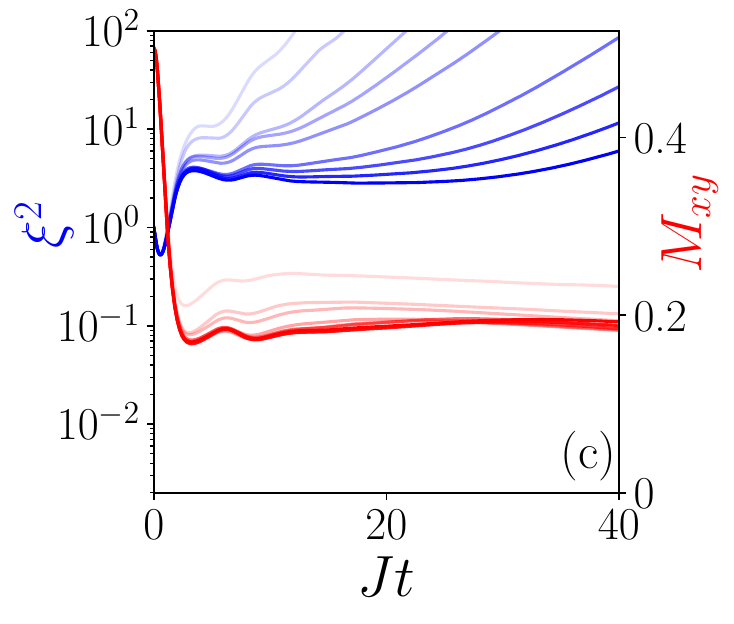}
\includegraphics[width=0.493\linewidth]{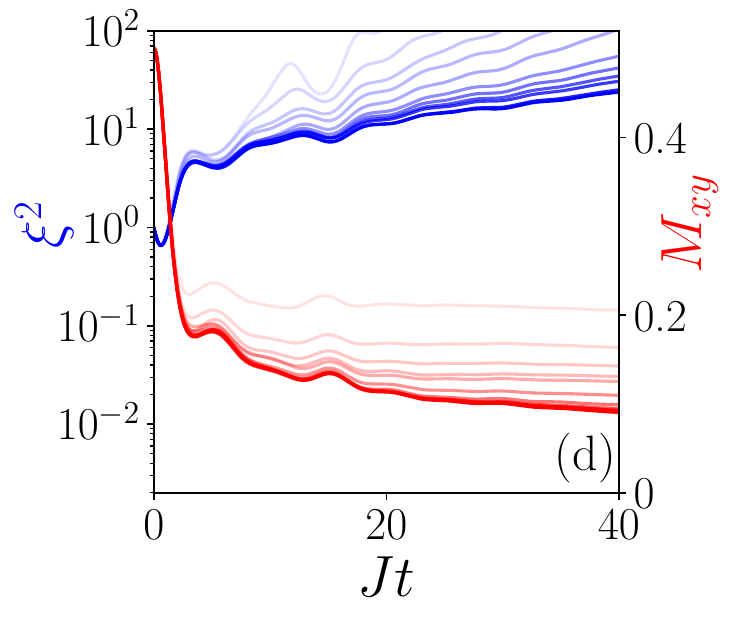}
    \caption{ Squeezing parameter $\xi^2$ (blue) and magnetization $M_{xy}$ (red) vs time for different $N$ values (faded lines) in the case of $\Delta = -1$.  The results correspond to vacancy probability (a) $p = 0$, (b) $p = 0.5$, (c) $p=0.75$, (d) $p = 0.85$. Opacity of lines scales with $\sqrt{N}$ over a system size range of approximately  $N \in \{10^2, 10^4 \}$. Results correspond to an average over 10 disorder realizations with 6400 dTWA samples for each system.}
    \label{fig:examples}
\end{figure}

\textit{Model.---}
We consider the spin-1/2 power-law interacting XXZ model 
\begin{align}
    H = - \sum_{i<j} \frac{J}{r_{ij}^3} (\hat{s}^x_i\hat{s}^x_j + \hat{s}^y_i\hat{s}^y_j + \Delta \hat{s}^z_i\hat{s}^z_j) \label{eq:XXZ}
\end{align}
with coupling constant $J$, anisotropy $\Delta$, distance $r_{ij}$ between spins $i$ and $j$, and spin operators $\hat{s}^{\alpha}_{i} = \hat{\sigma}_i^{\alpha}/2$ that obey the canonical commutation relations.  The spins lie in a square 2D lattice of length $L$ with lattice spacing $a$, which we set to unity, with each site occupied with probability $f = 1-p$, where $p$ is the vacancy probability. The average number of spins is therefore $N = fL^2$. We choose $1/r^3$ interactions as this is commonly realized in experiments \cite{Defenu2023}, including the recent NV experiment of Ref. \cite{Wu2025}. The 2D geometry eliminates the angular dependence of the dipolar interaction. To characterize the squeezing we use the spin squeezing parameter \cite{Wineland1992,Wineland1994} 
\begin{align}
  \xi^2 =  \frac{N {\rm min}_{\mathbf{n} \perp x} {\rm Var}[\mathbf{n}. \mathbf{\hat{S}}] }{\langle \hat{S}^x \rangle^2 }  
\end{align}
where $\hat{S}^{\alpha} = \sum_i \hat{s}^{\alpha}_i$ is the collective spin and $\mathbf{n}$ is the direction for which the variance is minimized in the plane perpendicular to the mean-spin vector, which we select as the x-direction. Scalable squeezing occurs when $\xi^2 \sim N^{-\nu}$, with $\nu \in \{0,1\}$, ranging from the standard quantum limit $(\nu = 0)$ to the Heisenberg limit ($\nu = 1$). In the disorder-free case, it has been shown that dynamics thermalizing to the easy-plane ferromagnetic ordered phase at late times display scaling $\xi^2_{\rm opt} \sim N^{-2/5}$ \cite{Block_2024}. However, in practice this may only be observable for extremely large systems, before which one observes one-axis-twisting (OAT) scaling $\xi^2_{\rm opt} \sim N^{-2/3}$ \cite{Block_2024}.  
Here, $\xi^2_{\rm opt}$ is defined as the minimum value of the spin squeezing parameter during the time evolution, which represents the point of maximal metrological utility. 
Since the disorder preserves U(1) symmetry we expect a similar picture to hold here, with the critical temperature $T_c$ lowering with increasing vacancy probability $p$.

We consider an initially x-aligned state $\ket{\psi(0)} = \prod_{i}^N \ket{+}_i$, which has a low enough energy density to ensure thermalization to the easy-plane ferromagnetic phase in the disorder-free case (for $-4 \lesssim \Delta \leq 1$) \cite{Block_2024}. We simulate Eq. (\ref{eq:XXZ}) using the semiclassical method known as the discrete truncated Wigner approximation (dTWA) \cite{Schachenmayer2015,Zhu2019}, which in the disorder-free case yields near-exact results for this system \cite{Muleady2023}.
The dTWA has previously been used for a range of disordered systems \cite{Kelly2021,Acevedo2017,Covey2018,Duha2024,Signoles2021,Schultzen2022}, with Ref. \cite{Duha2024} providing explicit results on the impact of finite filling fractions for the case of two-mode squeezing, suggesting a tolerance for disorder.  In the presence of strong disorder, this approximation is  not expected to capture the dynamics of strongly interacting spins \cite{Block_2024}.
In the Supplemental Material (SM) \cite{SM} we show that for $p \lesssim 0.7$ we observe near-quantitative agreement with a cluster variant of the method (cTWA) \cite{Braemer2024} over the relevant squeezing time-scales, in which the strongest interactions are treated exactly.

\begin{figure}[b]
    \centering
    \includegraphics[width=\linewidth]{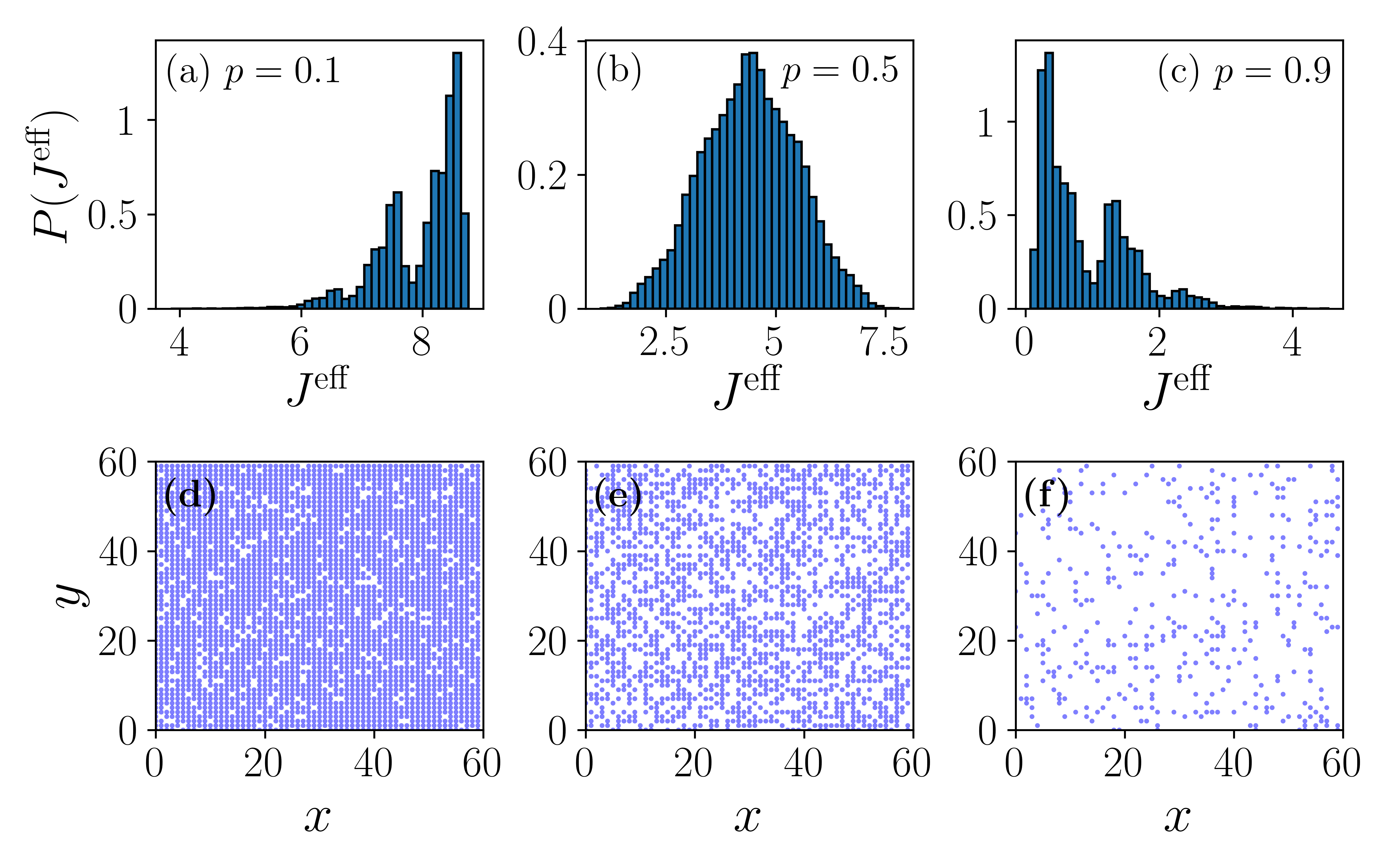}
    \caption{Distribution of effective interaction strengths $P(J^{\rm eff})$ for varying vacancy probability (a) $p=0.1$, (b) $p = 0.5$, (c) $p = 0.9$. (d)-(f) Corresponding spatial positions of spins (dark points) in the 2D plane for a single disorder realization. Results in (a)-(c) are obtained from  lattices with $N\sim \mathcal{O}(10^3)-\mathcal{O}(10^4)$, and are averaged over 25 disorder realizations.}
    \label{fig:distributions}
\end{figure}

\begin{figure}[t]
    \centering  
\includegraphics[width=\linewidth]{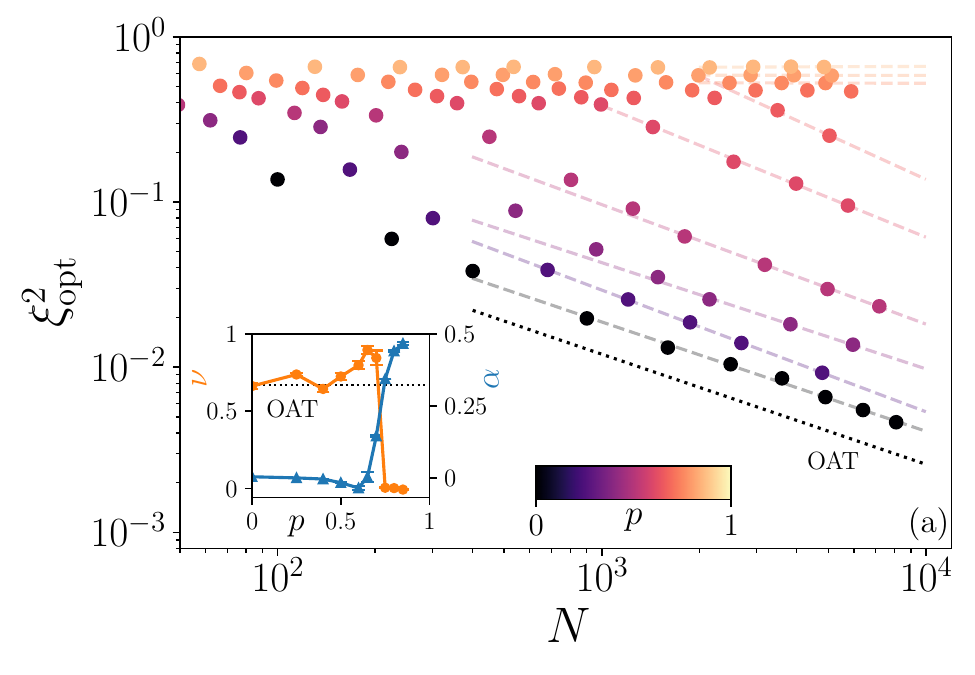}

\includegraphics[width=0.98\linewidth]{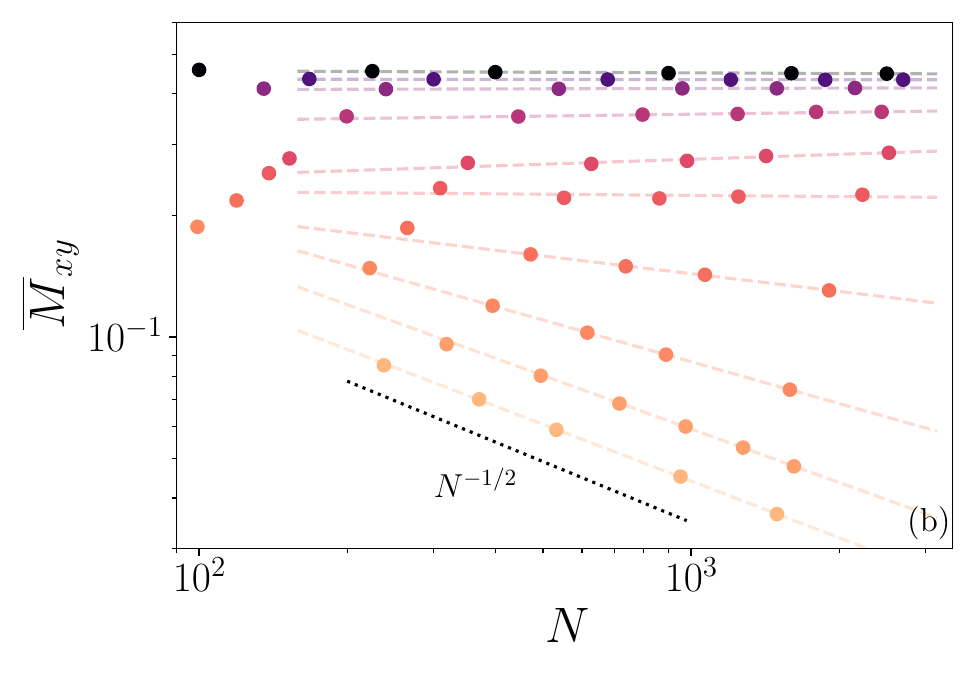}
    \caption{(a) Optimal squeezing parameter $\xi_{\rm opt}^2$ vs $N$  for a range of vacancy probabilities $p$, where the colors range from $p=0$ to $p=0.85$ (legend) and we have set $\Delta = -1$. The dotted line gives the scaling for OAT: $\xi_{\rm opt}^2\sim N^{-2/3}$. Inset: $\nu$ (circles) vs $p$ and $\alpha$ vs $p$ (triangles), extracted from fits to the data in panels (a) \cite{Footnote_xi2} and (b) respectively (dashed lines). 
    Error bars indicate uncertainty of the fit.  (b) Late-time magnetization  $\overline{M}_{xy}$ vs system size $N$ for different $p$ values [same legend as (a)].  
    Data in (a)[(b)] is obtained from an average over 10 (25) disorder realizations, with 6400 (1024) dTWA samples for each system. }
    \label{fig:scaling}
\end{figure}

\textit{Spin squeezing with disorder.---}
In Fig. \ref{fig:examples} we consider the case of $\Delta = -1,$ which is relevant for NV centers in diamond \cite{Wu2025}. We display results for the spin squeezing parameter $\xi^2$ (blue) and easy-plane magnetization $M_{ xy} = \sqrt{ \langle (\hat{S}^x)^2 + (\hat{S}^y)^2 \rangle}$ (red) vs time for different vacancy fractions $p$. Fig.~\ref{fig:examples}(a) shows the disorder-free case ($p=0$) for a range of system sizes (opacity), clearly illustrating that the minimum squeezing parameter $\xi^2_{\rm opt}$ decreases with system size, as expected for scalable squeezing, and observes similar scaling to the OAT model $\xi^2_{\rm opt} \sim N^{-2/3}$ [look ahead to Fig. \ref{fig:scaling}(a) inset].  The magnetization  quickly relaxes to a finite value, indicating  thermalization to the ordered phase. Fig.~\ref{fig:examples}(b) shows results for $p = 0.5$. While $\xi^2_{\rm opt}$ is not as small as the disorder-free case, it still demonstrates a decrease with system size similar to the OAT scaling. Fig. \ref{fig:examples}(c) shows the $p=0.75$ case. Here, the minimum occurs at early times on a scale set by the typical interaction strength and is not scalable. However, the second local minima does demonstrate scalable squeezing and for sufficiently large $N$ [well beyond the maximum accessible  $N \sim \mathcal{O}(10^4)$] will yield scalable squeezing. The time to reach this minimum also becomes large, which reflects the associated  critical slowing down near the finite temperature phase boundary.
Fig.~\ref{fig:examples}(d) shows $p = 0.85$, which does not exhibit  scalable squeezing. The magnetization  is also seen to decrease with increasing system size. The magnetization reaches a quasi-steady state at late times (not shown, examples in SM \cite{SM}), which we denote by $\overline{M}_{xy}$, and  decays as a power-law  with increasing $N$, $\overline{M}_{xy} \sim N^{-\alpha}$, indicating thermalization to the disordered phase in the thermodynamic limit. Within the disordered phase, and away from the phase boundary, we find $\overline{M}_{xy} \sim N^{-1/2}$, in keeping with analytic behavior in $L^{-1}$.

These results can be qualitatively understood from the distribution of the effective interaction strengths $P(J^{\rm eff})$, where $J^{\rm eff}_i = \sum_{j} J r_{ij}^{-3},$ shown in Fig.~\ref{fig:distributions} for a variety of $p$ values. For weak disorder $p=0.1$, Fig. \ref{fig:distributions}(a), it can be seen that the largest $J^{\rm eff}$ values are double that of the smallest values. 
In contrast, when $p = 0.9$ [Fig.~\ref{fig:distributions}(c)] this ratio becomes very large due to close pairs of spins having much stronger interactions than spins that are isolated in space -- see Fig. \ref{fig:distributions}(f) for the corresponding spatial distribution. For $p \approx 1,$ the typical spacing between spins far exceeds the lattice spacing and the distribution acquires a very heavy tail \cite{Wu2025}.   
Nevertheless, 
for  moderate disorder (such as $p = 0.5$ [Fig. \ref{fig:distributions}(e)]) 
all spins experience interactions of a comparable order of magnitude, which assists the persistence of collective squeezing behavior in Fig. \ref{fig:examples}.

Fig.~\ref{fig:scaling}(a) shows the optimal spin squeezing parameter $\xi^2_{\rm opt}$ vs $N$ for different $p$ values (colors). 
Following Ref. \cite{Block_2024}, we consider a nuanced definition of $\xi_{\rm opt}^2$ to eliminate the non-scalable minima  [as in Fig. \ref{fig:examples}(c)] \cite{Footnote_xi2}.
For small $p$ the results can be seen to track a power-law over at least an order of magnitude in $N$. Above a critical disorder strength $p=p_c$ we observe $\xi^2_{\rm opt} \sim N^0$ (const.). This is reflected in the inset, which shows the power-law $\nu$ vs $p$ extracted from a log-log fit to the data. Near the transition, $\nu$ values larger than the OAT result of $\nu = 2/3$ are likely to be finite-size effects:  even the disorder-free case shows a larger power-law for small $N$, only approaching OAT scaling at the larger system sizes we consider. 

Since scalable squeezing occurs for dynamics thermalizing to the ordered phase, an alternate approach is to use the late-time magnetization $\overline{M}_{xy}$ to diagnose the potential for scalable squeezing. In Fig.~\ref{fig:scaling}(b) we show $\overline{M}_{xy}$ vs $N$ for the same system. The data obeys a power-law $\overline{M}_{xy} \sim N^{-\alpha}$ for a range of $p$ values (dashed-lines). In the inset of Fig.~\ref{fig:scaling}(a), we plot both $\alpha$ and $\nu$, which take non-zero values in the disordered/ordered (scalable/non-scalable squeezing) phases respectively. While the inferred transition points are slightly offset, both indicate a transition in the range $p_c\approx 0.65-0.75.$ In the SM \cite{SM} we provide similar data for an example with more points in the disordered phase ($\Delta = -2$), for which  an extremely clean power-law scaling is visible.

\begin{figure}[t]
    \centering
    \includegraphics[width=\linewidth]{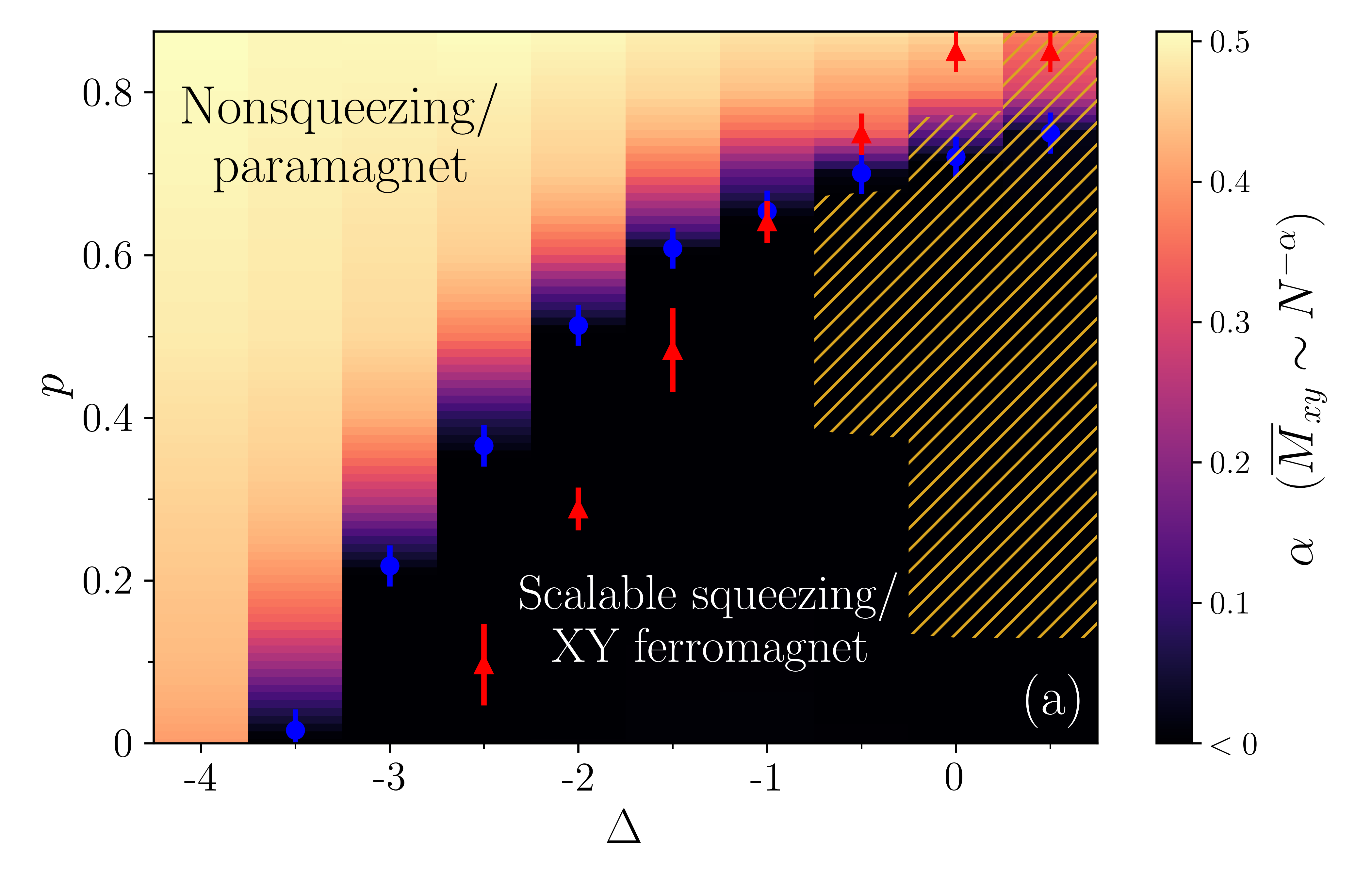}

     \includegraphics[width=\linewidth]{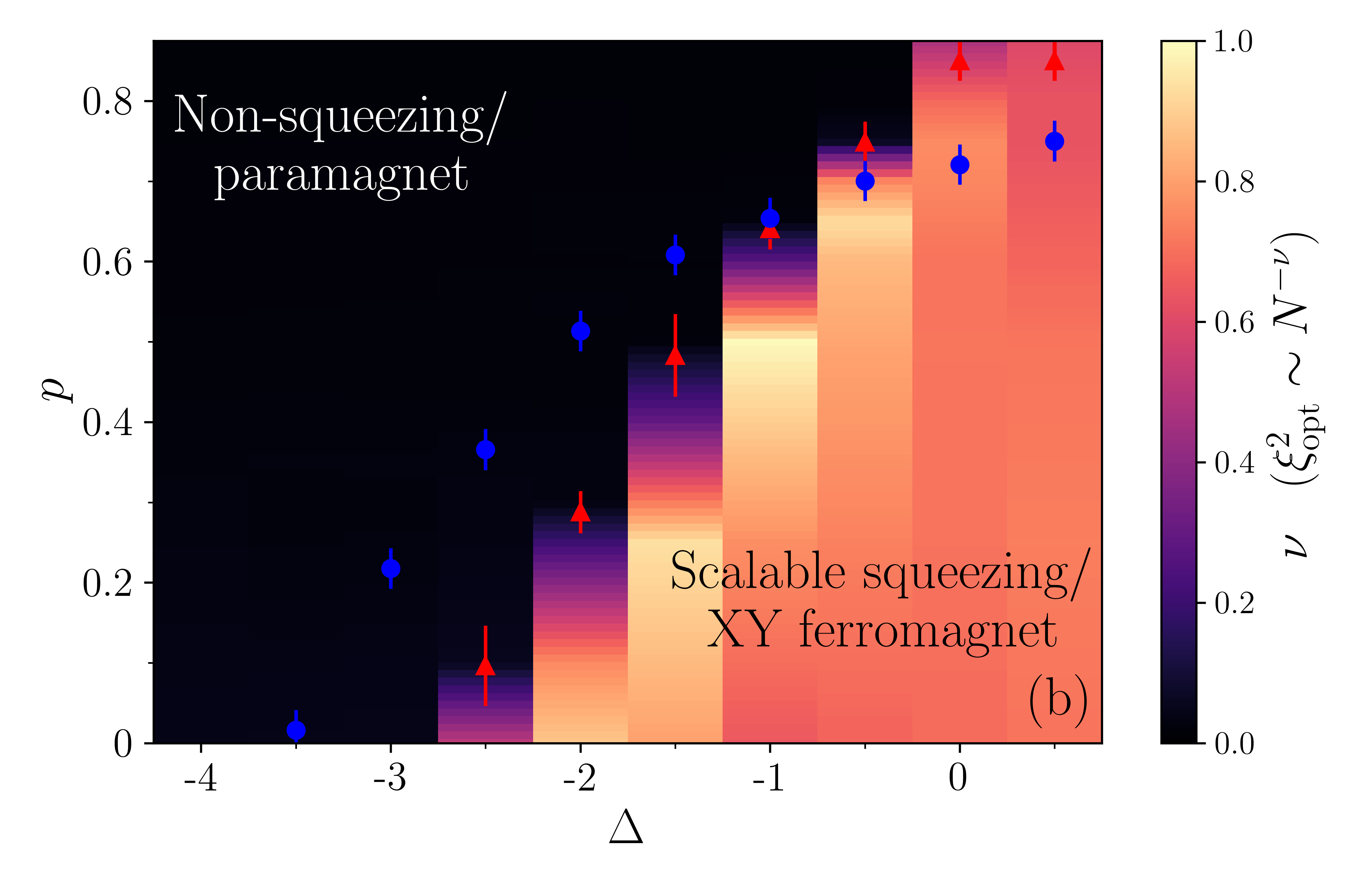}
    \caption{(a) Scaling exponent $\alpha$ (color) as a function of vacancy fraction $p$ and anisotropy $\Delta$, where $\overline{M}_{xy} \sim N^{-\alpha}$. (b) Squeezing exponent $\nu$ (color) as a function of vacancy fraction $p$ and anisotropy $\Delta$, where $\xi^2_{\rm opt} \sim N^{-\nu}.$   Blue dots correspond to estimates of the phase boundary from $\alpha$ [data in (a)], while red triangles correspond to the estimate from $\nu$ [data in (b)]. Results in (a) [(b)] correspond to an average of 25 disorder realizations and 1024 (12800) dTWA samples for each system, with the largest system sizes typically in the range $N \in \{1500,3000\}$. Error bars include sampling uncertainty and discreteness of simulated $p$-grid \cite{SM}.  Color data in the y-direction is linearly interpolated. For $\Delta =0$ and $\Delta = 0.5$, red triangles represent a lower bound of $p_c =0.85$ since we see scalable squeezing ($\nu >0$) for the entire  $p-$grid. Hatched region in (a) indicates simulations where the magnetization has not fully converged to the steady-state \cite{SM}}, in which data is taken for the largest times available.
    \label{fig:phase_diagram}
\end{figure}

\textit{Spin squeezing phase diagram.---} We now look to establish a spin squeezing phase diagram in the $p-\Delta$ plane. Fig. \ref{fig:phase_diagram}(a) shows the finite-size magnetization exponent $\alpha$, while Fig.~\ref{fig:phase_diagram}(b) shows the spin squeezing exponent $\nu$, where $\xi^2 \sim N^{-\nu}$. Both diagnostics predict a large region of scalable squeezing in the $p-\Delta$ plane, with increased robustness to disorder for the larger $\Delta$ values. We refrain from showing results in the region $0.5<\Delta <1,$ where the ordered phase persists but the dynamics are extremely slow and spin squeezing occurs only at inaccessibly late times \cite{SM}.  
The slow dynamics are primarily due to the proximity to $\Delta = 1$, where the initially x-aligned state is a ground-state (for all $p$), and hence no dynamics occur. In the disorder-free case, rotor-spin-wave theory predicts that the squeezing time-scale diverges as $(1-\Delta)^{-1}$  \cite{Roscilde2022,Roscilde2023}. Near the phase boundary the slow dynamics is further exacerbated by critical slowing down; in the ordered phase we find $t_{\rm opt} \sim N^{\mu}|p-p_c|^{-\gamma}$ for $\Delta = -1$, where $\mu \approx 1/3$ (the OAT result) and $\gamma \simeq 0.91(5)$ \cite{SM}. 
Given the invariance of the ground-state to disorder at $\Delta = 1$, it is natural to expect that the low-energy sector near this point retains a ferromagnet-like tower-of-states structure that is weakly affected by disorder. This in turn supports the persistence of scalable squeezing  over a broad range of $p$, as the collective dynamics remain OAT-like up to finite-temperature corrections \cite{Roscilde2022,Roscilde2023,Block_2024}.
Even for $\Delta = 0$ we observe scalable squeezing for the entire simulated region $p \in \{0,0.85\}$.
We do not present data for $p > 0.85$, as dTWA is not expected to provide a good approximation for a heavy tailed interaction distribution \cite{SM}. The phase diagram is suggestive that this region may display squeezing for $\Delta \gtrsim 0$.

While $\alpha$ and $\nu$ predict a qualitatively similar phase boundary, there are clear differences near the transition (blue dots vs red triangles). This is attributed to the absence of scalable squeezing in this region at the accessible system sizes, in analogy to the results in Fig. \ref{fig:examples}(c). 
We anticipate that the phase diagram in Fig.~\ref{fig:phase_diagram}(a) is more accurate for $\Delta \leq -1$ as the magnetization data is less affected by finite-size corrections (beyond the $N^{-\alpha}$ scaling). However, for $\Delta \geq -0.5$ the magnetization typically does not always fully converge to the steady state over accessible timescales, relaxing over timescales far beyond the squeezing time  \cite{SM}; in this region we take the largest times available (hatched region). Nevertheless, comparison with Fig.~\ref{fig:phase_diagram}(b) shows scalable squeezing for the majority of these parameters.  
Interestingly,  the magnetization data suggest a disordered phase for $\Delta = 0$ and $p=0.85,$  in conflict with the scalable squeezing observed in Fig.~\ref{fig:phase_diagram}(b). However,
comparisons with cluster (cTWA) calculations suggest results at the most extreme $p$ values we consider are at best qualitative \cite{SM}.

\textit{Discussion.---}
The results in Fig.~\ref{fig:phase_diagram} illustrate why the recent NV experiment in Ref. \cite{Wu2025} did not observe scalable squeezing: the strong disorder ($p$ close to unity) far exceeded our best estimate of the critical value $p_c = 0.75(3)$ for $\Delta = -1$, extracted from the scaling analysis of $\xi^2_{\rm opt}$ in Fig.~\ref{fig:scaling}(a).

A promising way to make scalable spin squeezing more realistic in defect ensembles is to engineer a near-ordered sensor lattice through controlled defect creation. One practical route is to use electron-beam lithography to define a nanoscale patterned implantation mask, then perform nitrogen ion implantation and thermal annealing to create NV centers. NV creation at each site is inherently stochastic and can be approximated as a Poisson process. The implantation dose can be adjusted to target an average of about one NV per lattice site. If the created NVs stay near the center of each site, then sites with multiple spins can be removed from their strong dipolar interaction via frequency resolved shelving or adiabatic depolarization, as outlined in Ref. \cite{Wu2025}. Under this assumption, the ``useful'' sites are the ones with exactly one NV, while both empty sites and multi-occupied sites effectively behave like vacancies/inactive sites. With a Poisson mean of one, the probability of getting exactly one NV at a site is $e^{-1}\approx 0.37$, with effective vacancy probability $p=1-0.37=0.63$. For $\Delta = -1$ this is expected to be in the scalable squeezing phase [Fig.~\ref{fig:phase_diagram}~(a)]. 

An intriguing alternative experimental platform is the boron vacancy (V$_\mathrm{B}$) center in hexagonal boron nitride (hBN)~\cite{gottscholl2020initialization,gong2024isotope, biswas2025quantum}. Compared to NV in diamond, V$_\mathrm{B}$ defects may offer a key advantage for scalable squeezing: the defect is structurally simpler (a single missing boron atom), and the two-dimensional host makes it possible to create and visualize individual defects at the nanoscale using scanning transmission electron microscopy (STEM)~\cite{kikkawa2025observation}. If defect placement can be made deterministic, this approach could overcome the stochastic Poisson statistics and enable a near-ordered defect lattice with an effective vacancy fraction well below 0.63, which would be even more favorable for accessing the scalable squeezing regime.

Our results also suggest that only large lattices may realize a practical squeezing advantage close to the phase boundary. This is irrelevant for NVs and other solid-state platforms with typical atom numbers of $N \sim \mathcal{O}(10^{12}),$ but should not be neglected in quantum gas experiments with arrays of size $N \sim \mathcal{O}(10^2-10^5)$ spins. Due to the disorder averaging, we are unable to reach the system sizes realized in Ref. \cite{Block_2024} for the disorder free case, which sees a cross-over to $\nu=2/5$ for $N > \mathcal{O}(10^{4})$. Since these atom numbers are experimentally relevant, future  work should interrogate this further for the disordered case. 

Finally, our results lend weight to the suggestion of Ref. \cite{Wu2025} that  Floquet engineering techniques \cite{choi2020robust} be employed to engineer $\Delta$ values deep in the ordered phase, where we find squeezing to be highly robust to disorder. Operating close to the point $\Delta = 1$ may require careful consideration of system coherence times, due to the long evolution required to reach the squeezing minima. 

Our work establishes the robustness of scalable squeezing to disorder in power-law interacting XXZ models. This is relevant for a large range of quantum simulation platforms, and provides a crucial step towards realizing quantum devices that yield a practical metrological advantage via spin squeezing. We also explain the absence of scalable squeezing in a recent NV experiment \cite{Wu2025}, provide potential strategies for disorder reduction, and highlight regimes with the best disorder tolerance.   \\

\textit{Acknowledgments.---}
We acknowledge support from the NSF
through awards OMR-2228725 (S.E.B. and M.H.K.) and DMR-1945529  (M.H.K.), and thank the
High Performance Computing facility at The University
of Texas at Dallas (HPC@UTD) for providing computational resources (S.E.B. and B.G.). C.Zu. acknowledges support from the NSF under Grant No. 2514391. C.Zh. acknowledges support from Air Force Office of Scientific Research under Grant No. FA9550-2010220 and from NSF OSI-2503230. B.G. acknowledges funding in part by the Austrian Science Fund (FWF) Grant-DOI: 10.55776/PAT3563424.

During the preparation of this manuscript, a pre-print \cite{Kaplan2025} appeared also demonstrating the effects of disorder on spin squeezing for the power-law interacting XXZ model on the 2D lattice. Our results support the conclusions of that work. They obtain the finite-temperature phase diagram via quantum Monte Carlo simulations, obtaining results that are qualitatively similar, with possible quantitative agreement with our results for intermediate $\Delta$ values.


%

\end{document}


\title{Supplemental Material for ``Scalable Spin Squeezing in Power-Law Interacting XXZ Models with Disorder"}

\author{Samuel E. Begg}
\affiliation{Department of Physics, The University of Texas at Dallas, Richardson, Texas 75080, USA}
\author{Bishal K. Ghosh}
\affiliation{Department of Physics, The University of Texas at Dallas, Richardson, Texas 75080, USA}
\affiliation{Institute of Physics, University of Graz, Universitätsplatz 5, 8010 Graz, Austria}
\author{Chong Zu}
\author{Chuanwei Zhang}
\affiliation{Department of Physics, Washington University, St. Louis, Missouri 63130, USA}
\affiliation{Center for Quantum Leaps, Washington University, St. Louis, Missouri 63130, USA}
\author{Michael Kolodrubetz}
\affiliation{Department of Physics, The University of Texas at Dallas, Richardson, Texas 75080, USA}

\date{\today}

\maketitle

\begin{figure*}
    \centering
 \includegraphics[width=0.39\linewidth]{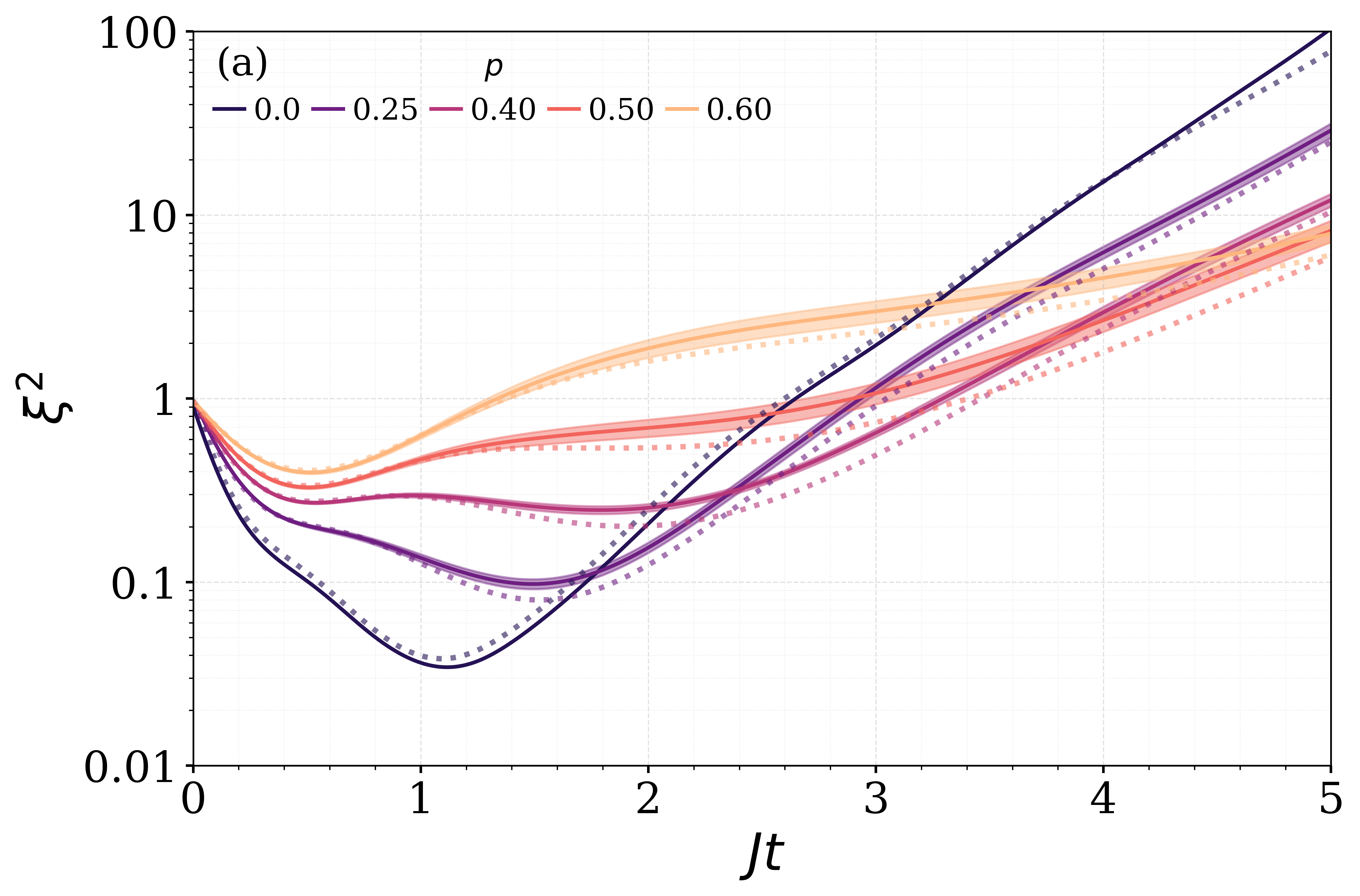}   \includegraphics[width=0.38\linewidth]{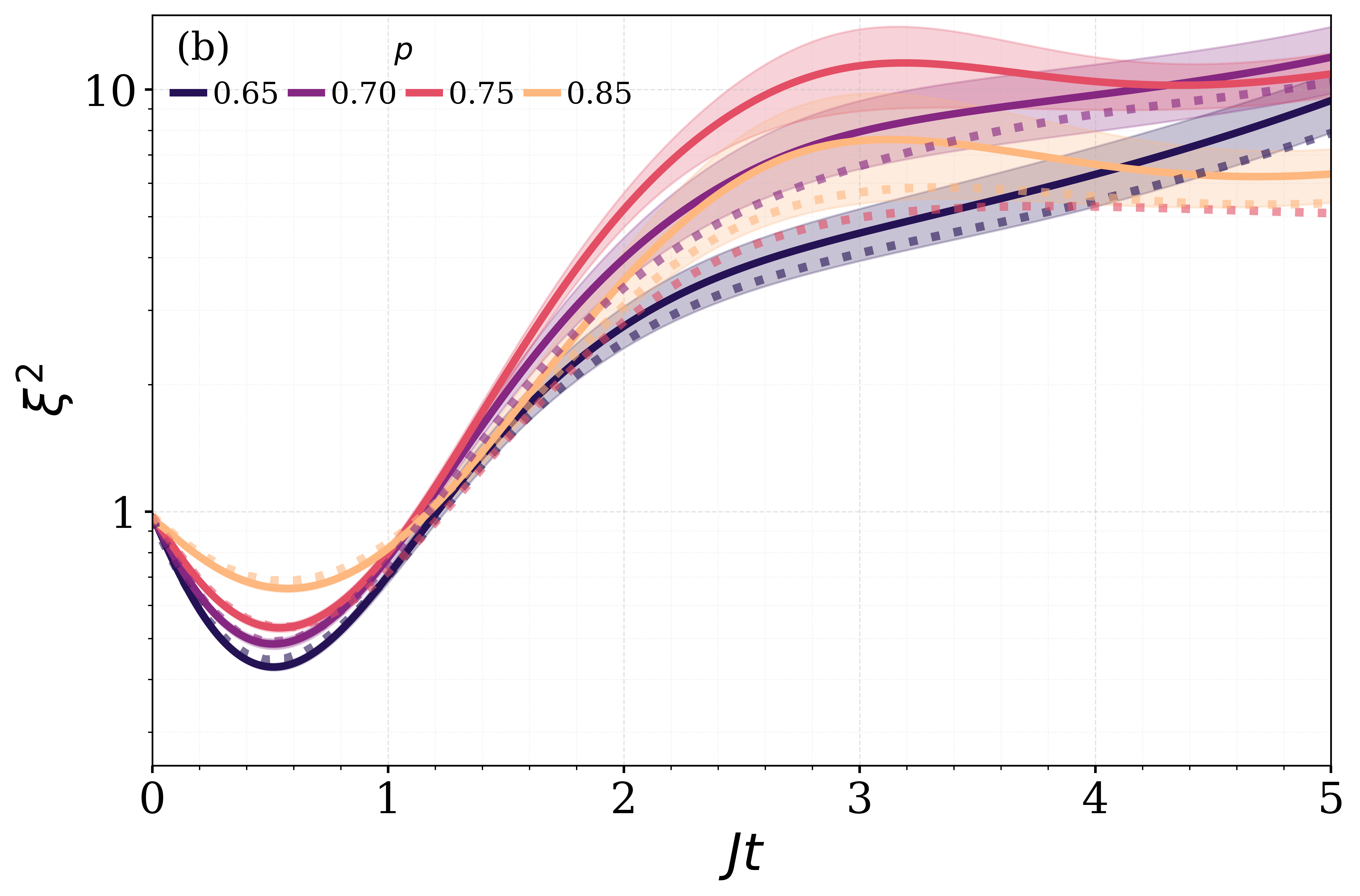}
 
\includegraphics[width=0.39\linewidth]{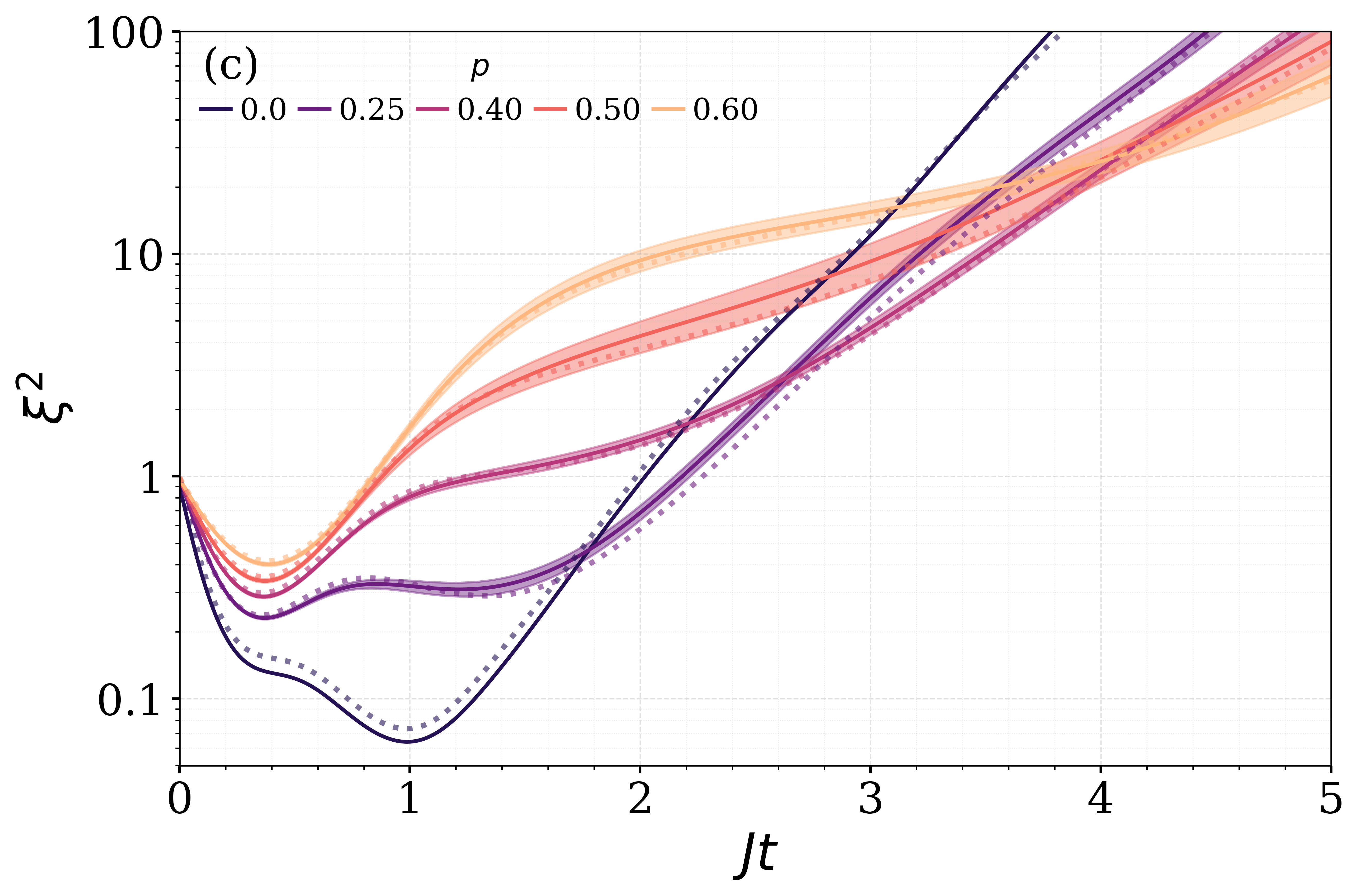}
 \includegraphics[width=0.39\linewidth]{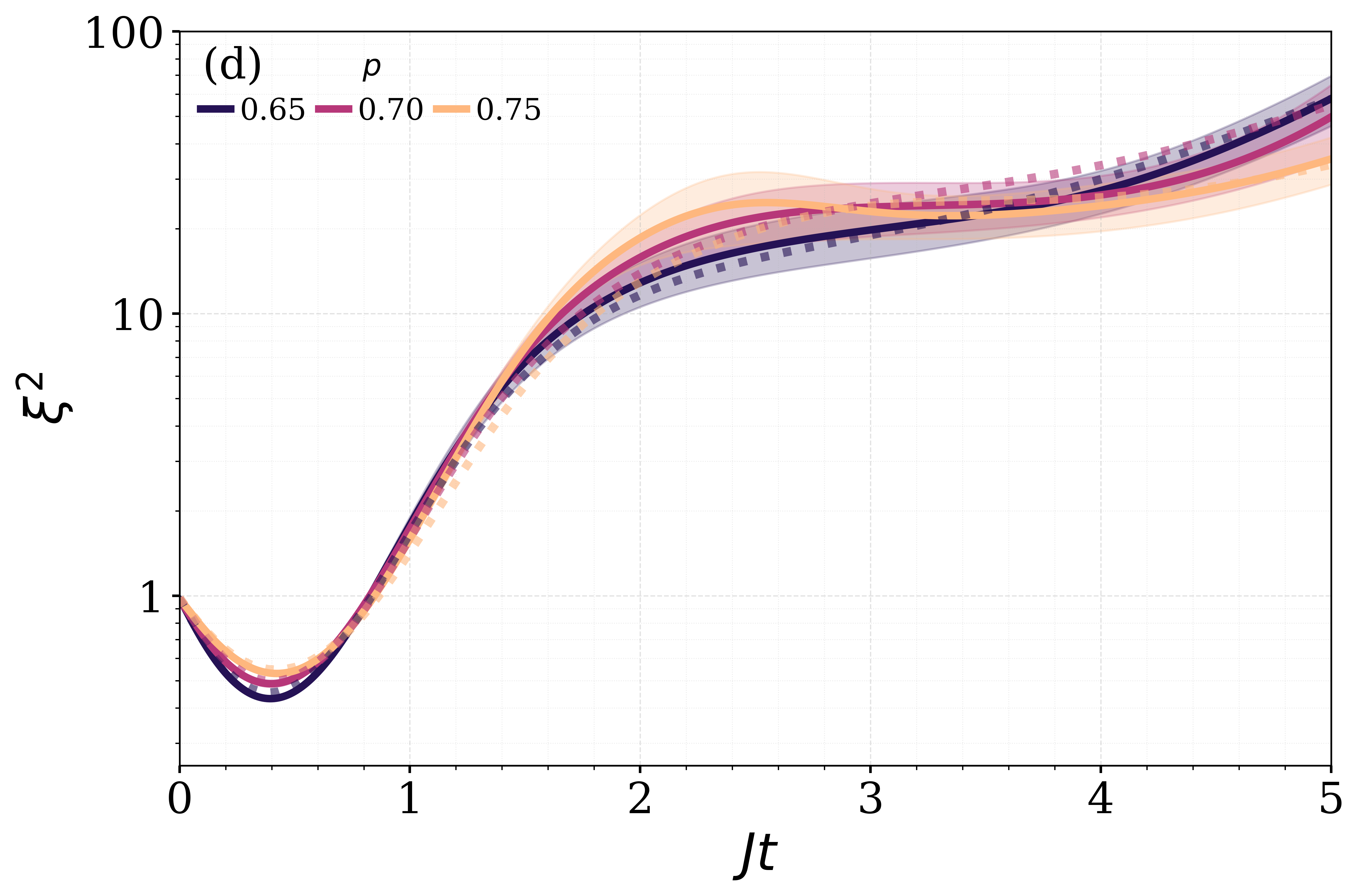} 
 
  \includegraphics[width=0.39\linewidth]{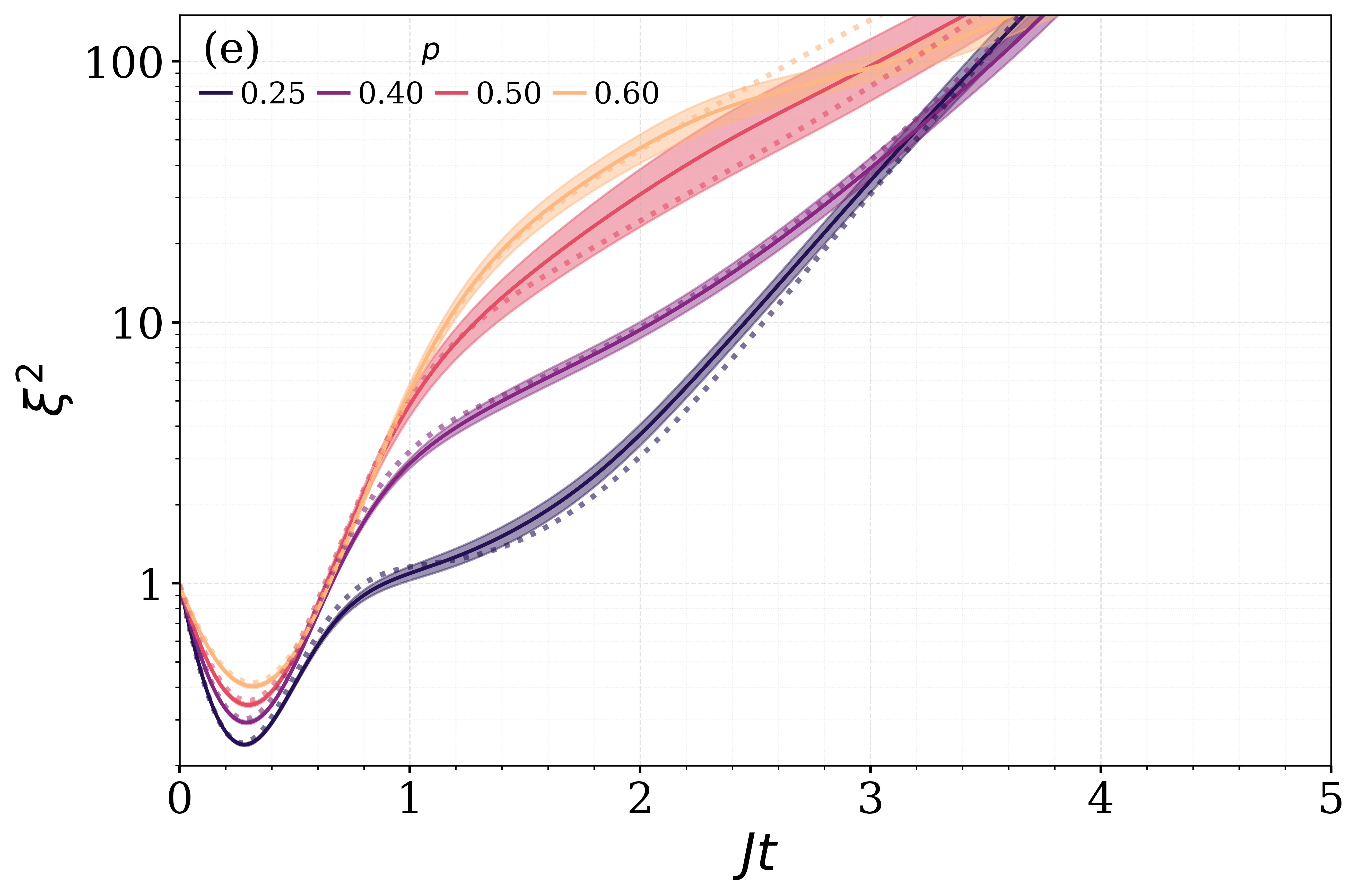}
  \includegraphics[width=0.39\linewidth]{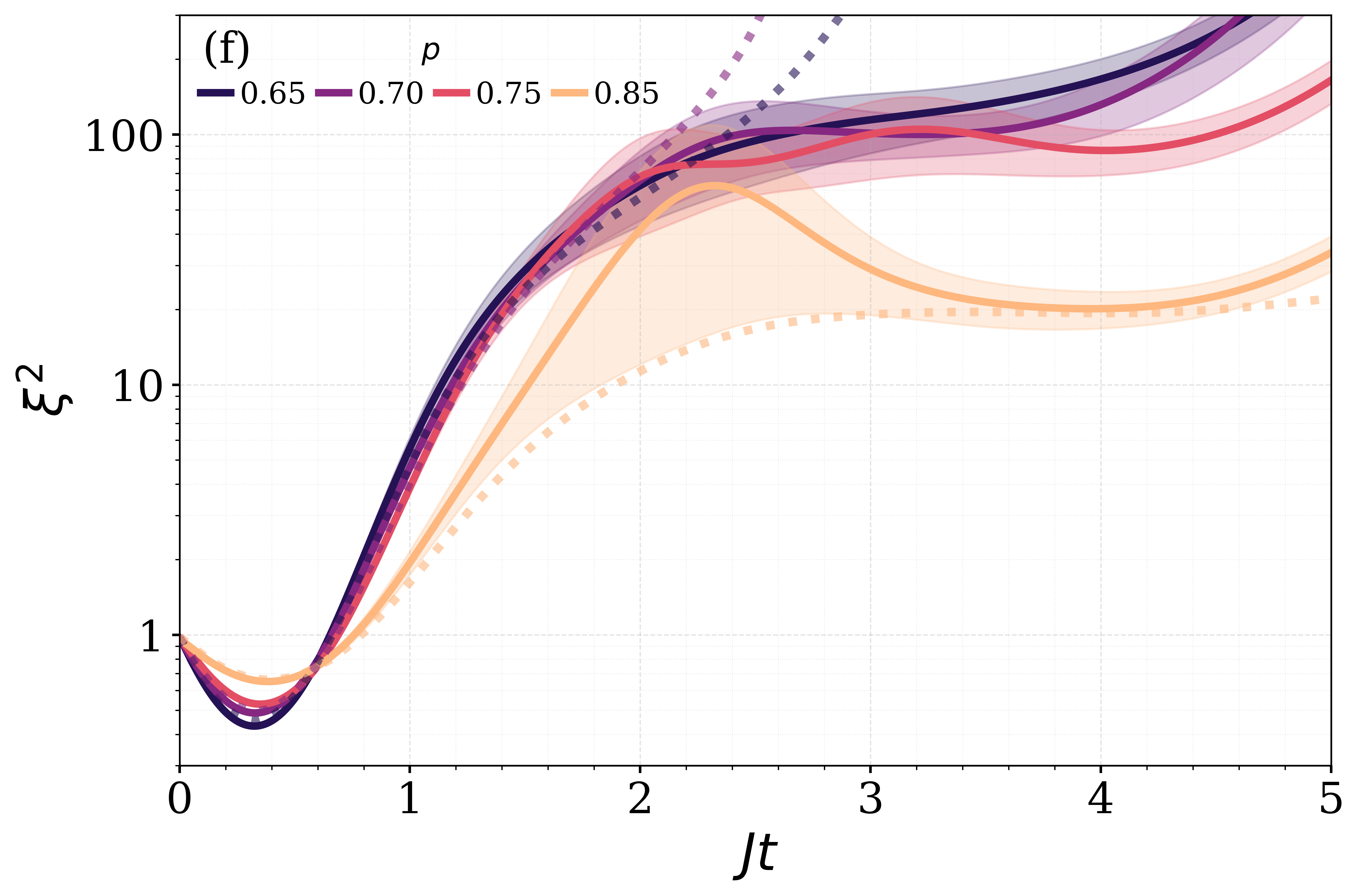}
    \caption{(a) Spin squeezing parameter $\xi^2$ vs time  for varying $p \in [0,0.65]$, setting $\Delta = -1.0$. (b) Same but for higher values $p \in [0.7,0.95]$, separated from (a) for ease of visibility. Solid lines indicate results from cTWA while dotted lines are those obtained with dTWA. Shading indicates disorder sampling error for cTWA. (c) and (d) show results for $\Delta = -1.5$, while (e) and (f) show results for $\Delta = -2$ . In all cases the data corresponds to fixed $L=20$ [$N=(1-p)L^2$].  For cTWA we take 10 disorder realizations with 1000 cTWA samples. For dTWA, in (a) and (b) we use 10 disorder realizations and 6400 dTWA samples, while (c)-(f) have 25 disorder realizations and 12800 dTWA samples.}
    \label{fig:cTWA_test}
\end{figure*}

\begin{figure*}
    \centering
    \includegraphics[width=0.325\linewidth]{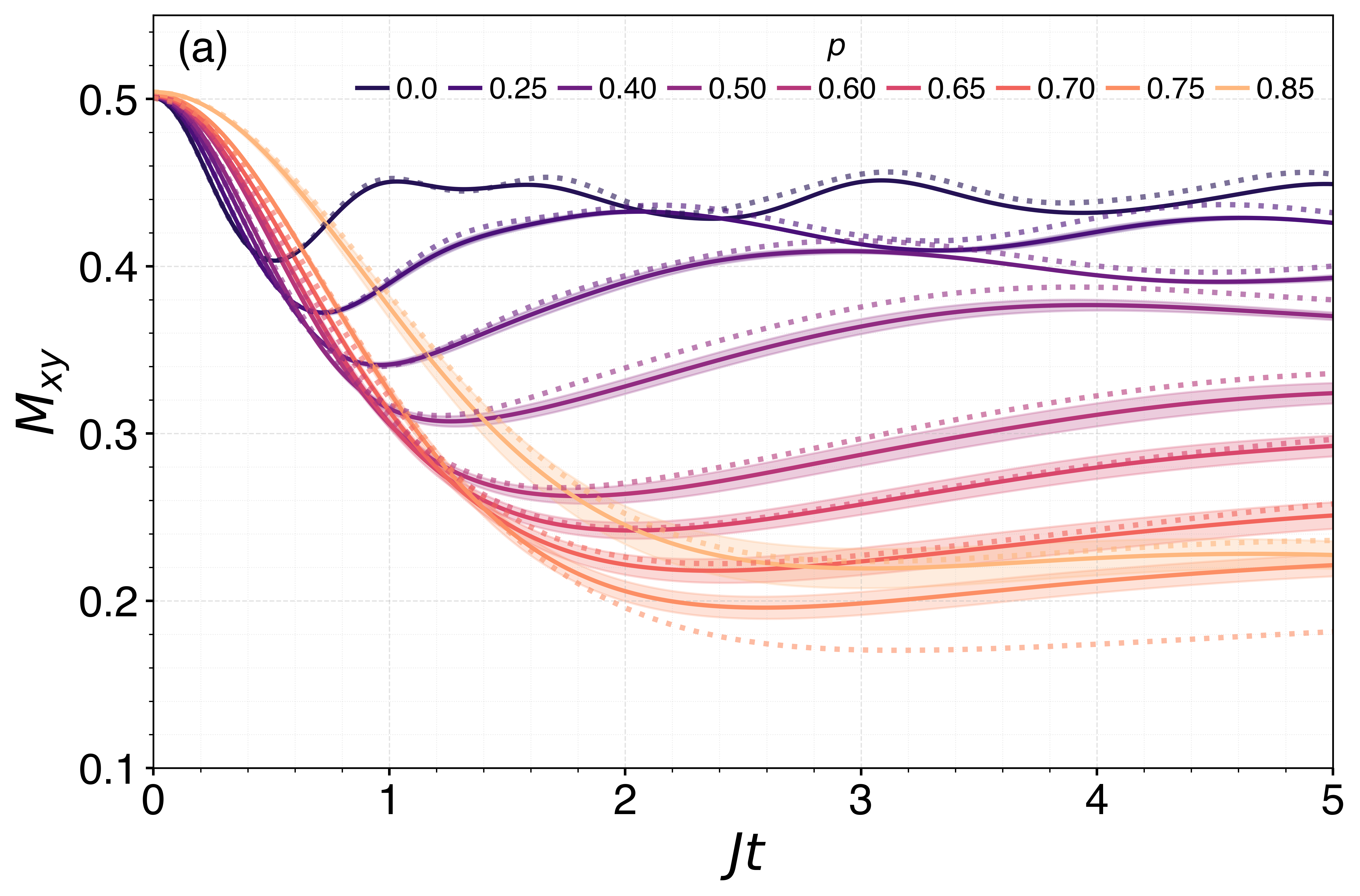}
   \includegraphics[width=0.325\linewidth]{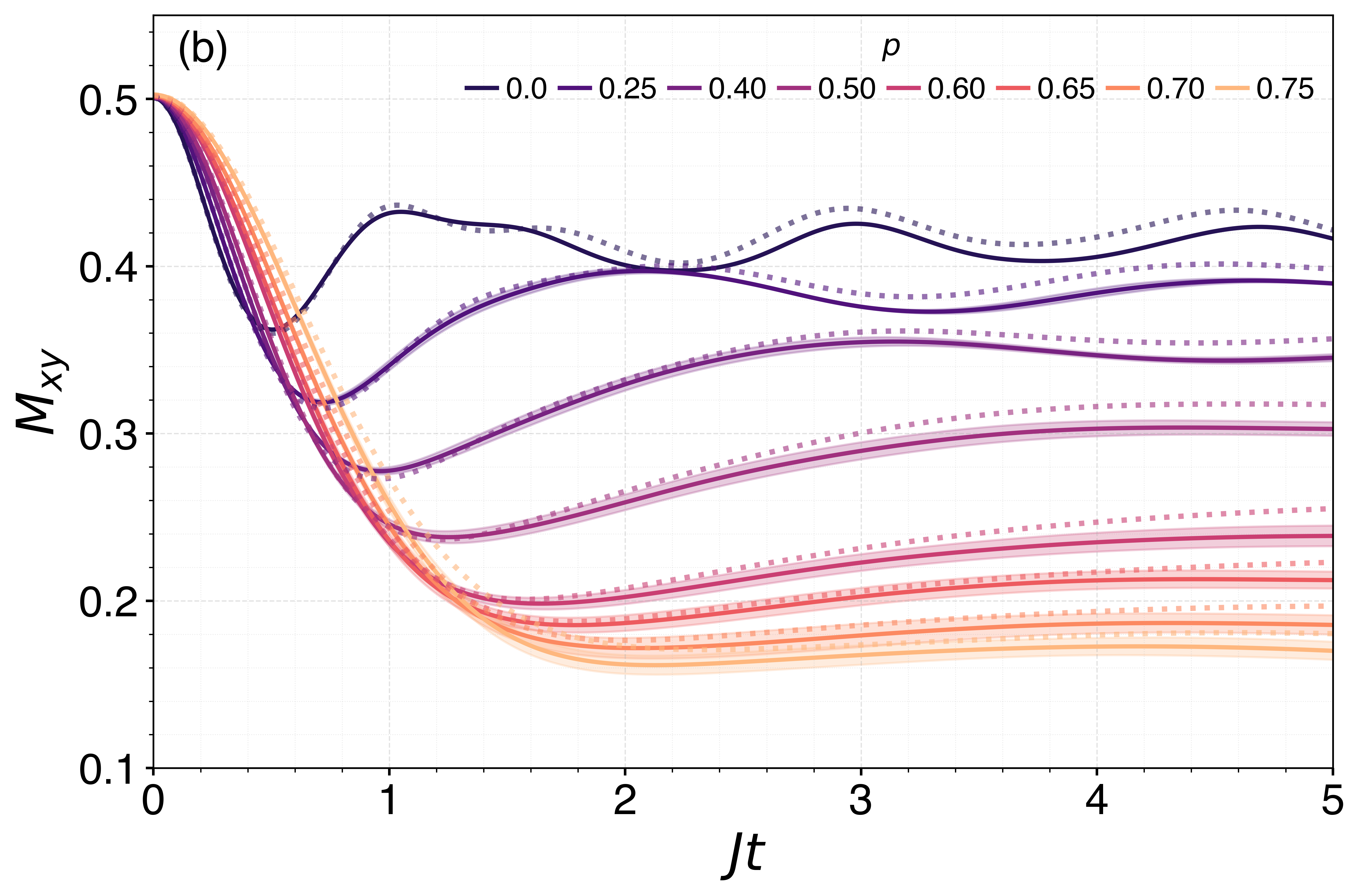}
    \includegraphics[width=0.325\linewidth]{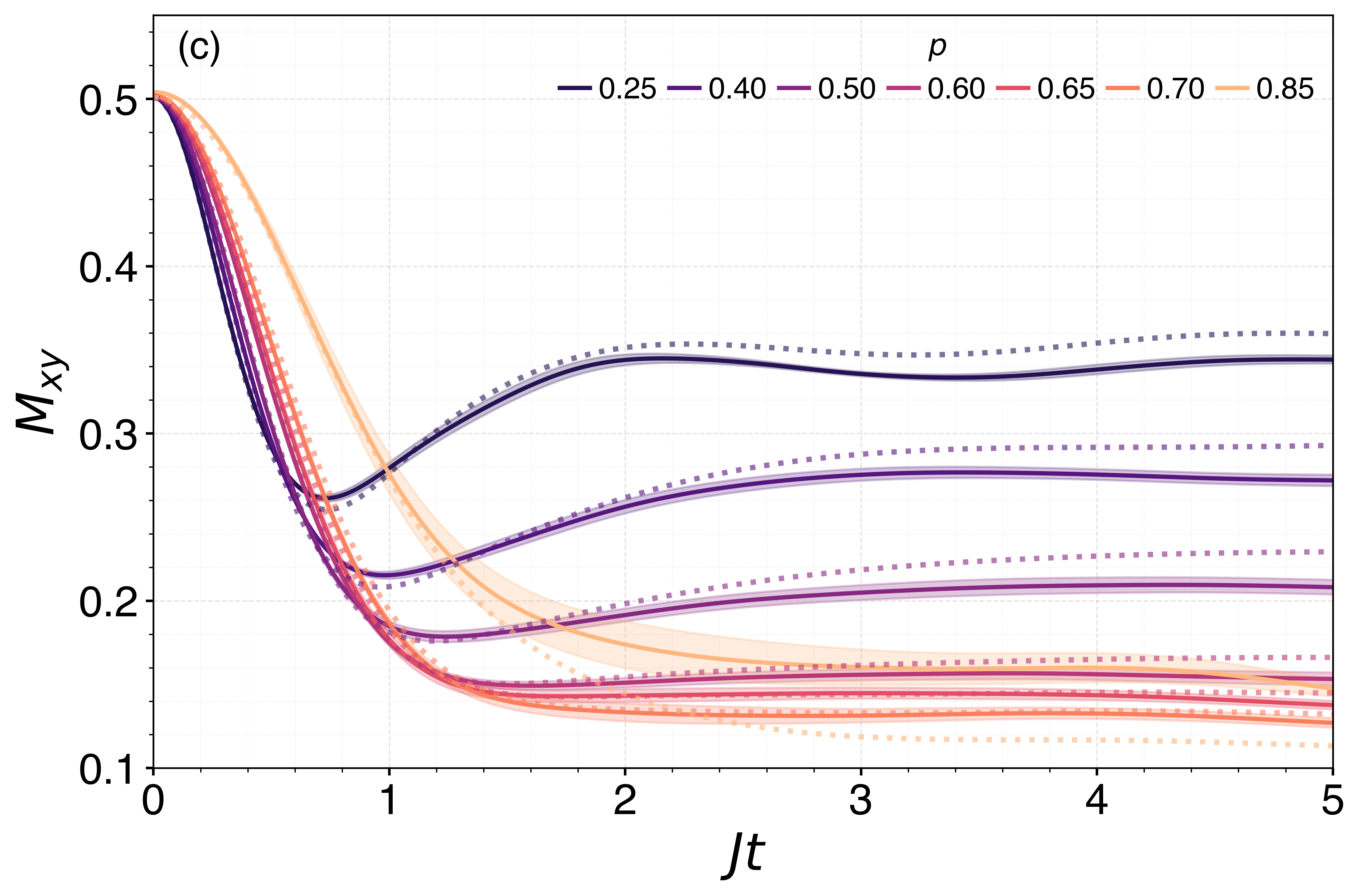}
    \caption{(a) Magnetization $M_{xy}$ vs time  for a different vacancy probabilities $p$ (legend), setting $\Delta = -1.0$. Solid lines indicate results from cTWA while dotted lines are those obtained with dTWA. Shading indicates disorder sampling error for cTWA. Panel (b)[(c)] shows similar results for $\Delta = -1.5$ ($\Delta=-2$). In all cases the data corresponds to fixed $L=20$ [$N=(1-p)L^2$]. For cTWA we take 10 disorder realizations with 1000 cTWA samples. For dTWA, in (a)  we use 10 disorder realizations and 6400 dTWA samples, while (b) and (c) correspond to 25 disorder realizations and 12800 dTWA samples.}
    \label{fig:cTWA_test_mag}
\end{figure*}

\section{Comparisons with the Cluster Truncated Wigner Approximation}
In this section we compare the results of the discrete truncated Wigner approximation (dTWA) \cite{Schachenmayer2015} used in the main text, with a cluster variant of the approach (cTWA) \cite{Braemer2024}. In the cluster approach, the dynamics within a cluster of spins is exact, while inter-cluster couplings are treated within the dTWA approximation. Here, we consider clusters of two spins, which allows us to capture the strongest interacting `dimer' pairs exactly, as discussed in Ref. \cite{Block_2024}. In order to sort all spins into pairs, we follow the prescription in Ref. \cite{Braemer2024}, starting by finding the strongest interacting pair of spins ranked by their interaction $J_{ij}$. We then consider the strongest interacting pair of the remaining spins, and proceed iteratively until all spins are paired. 

Fig. \ref{fig:cTWA_test}(a) shows the results for $\Delta = -1$ and $L = 20$, for vacancy probabilities in the range $p\in [0,0.65]$. It can be seen that the dTWA results typically lie just outside the disorder sampling uncertainty of the cTWA data (shaded regions). The small correction for cTWA suggests that dTWA provides near-quantitative accuracy over this range of $p$. Furthermore, these estimates from small system sizes correspond to a conservative scenario, since dTWA accuracy typically improves with system size.  Fig. \ref{fig:cTWA_test}(b) considers larger $p$ values. While $p=0.7$ remains in good agreement, a large deviation is noted for the case of $p=0.75$. This reflects the importance of short-range, strongly interacting pairs, which are treated exactly in cTWA but only approximately in dTWA. For $p \geq 0.8$, while both methods show similar results, we believe that they are no longer quantitatively accurate due to the emerging broad distribution of interactions -- see Fig. \ref{fig:distributions_large} for $P(J^{\rm eff})$ distributions over a range of $p$ values. 

Fig.~\ref{fig:cTWA_test}(c) and (d) exhibit a qualitatively similar trend for $\Delta = -1.5$ over the range of $p$ values displayed. Similarly, Fig.~\ref{fig:cTWA_test}(e) and (f) show comparable results for $\Delta = -2$. 
Discrepancies between dTWA and cTWA can  become more pronounced when $\xi^2 \gg 1,$ however, this regime lies outside the relevant squeezing window and does not affect the extracted $\xi^2_{\rm opt}$. In particular, when evaluating $\xi^2$ we always see good agreement for $p \lesssim 0.7$ over the time scales relevant for extracting $\xi^2_{\rm opt}$.

In Fig.~\ref{fig:cTWA_test_mag} we show a comparison between the two approaches for the magnetization vs time. For (a)~$\Delta = -1$ the results exhibit near-quantitative accuracy up to $p = 0.75$ over the time window shown. Similar behavior is observed in panels (b)~$\Delta =-1.5 $ and (c)~$\Delta =-2$, which show agreement between the methods at early times for the displayed values $p\lesssim 0.7$, with noticeable discrepancies at late times, but retaining qualitative agreement in the overall relaxation behavior.

\begin{figure}
    \centering
    \includegraphics[width=\linewidth]{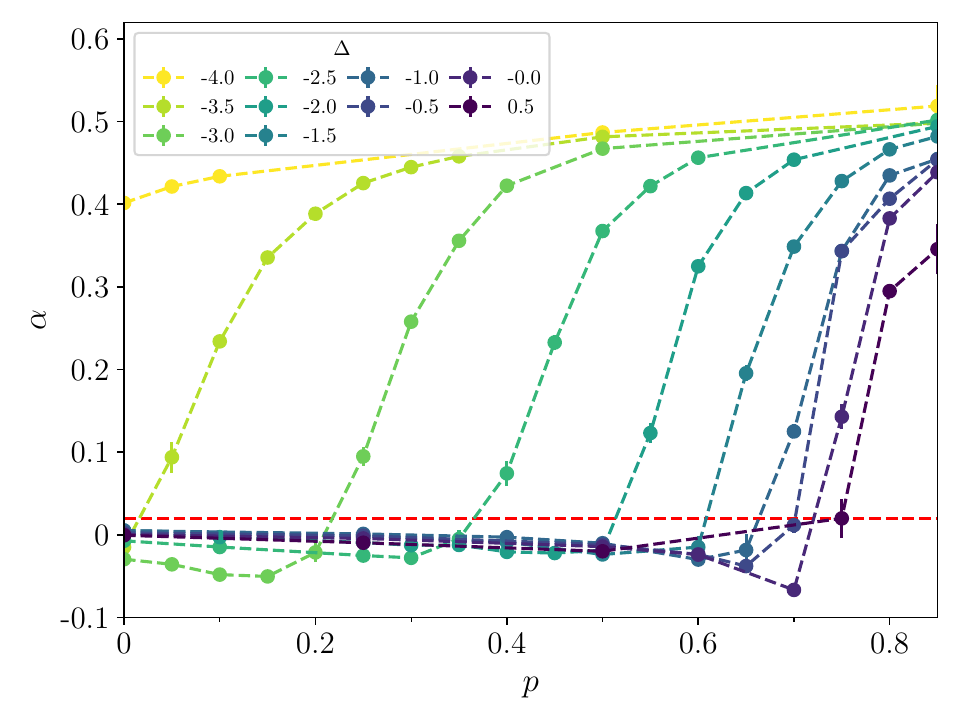}
 \includegraphics[width=\linewidth]{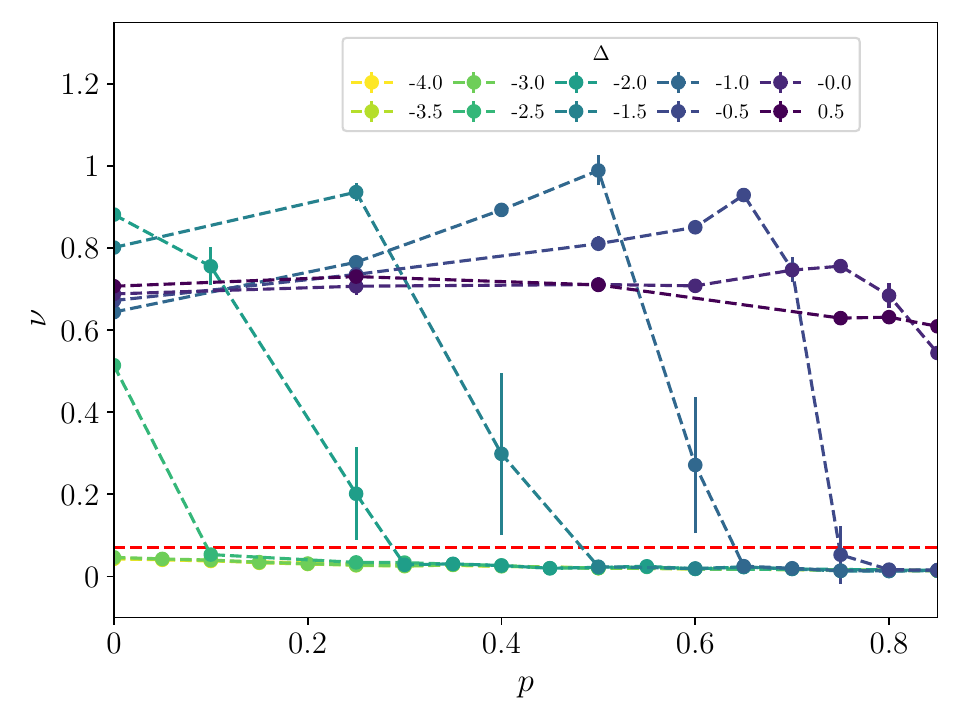}
    \caption{(a) $\alpha$ vs $p$ for a range of $\Delta$ values (legend). (b)  $\nu$ vs $p$ for a range of $\Delta$ values (legend). In both plots error bars indicate the uncertainty, though these are small and hence not always visible in (a).}
    \label{fig:data_phase}
\end{figure}

\section{Numerical Error analysis}
To estimate the error on our numerical results we need to consider both  dTWA sampling error and  disorder sampling error. For a given observable $\hat{\mathcal{O}}$ and a single disorder realization, we calculate the mean $\mathcal{O}_{\rm dtwa}$ and the standard error of the mean $\sigma_{{\rm dtwa}}$ for the dTWA samples. The full error after disorder averaging contains a direct contribution given by $\sigma^2_{\rm dis} =  \frac{1}{\mathcal{N}(\mathcal{N}-1)} \sum_s^{\mathcal{N}} (\mathcal{O}_{\rm dtwa,s} - \overline{\mathcal{O}_{\rm dtwa}})^2 $, where the sum is over the sample index $s$ and the bar denotes the disorder average over $\mathcal{N}$ samples. This is combined with the aforementioned dTWA error to yield the total standard error of the mean
\begin{align}
    \sigma_{\rm SEM} = \sqrt{\sigma_{\rm dis}^2 + \overline{\sigma^2_{{\rm dtwa}}}/\mathcal{N}}  .
\end{align}
In the case of $\xi^2$, the error is estimated via standard error propagation using the individual errors for ${\rm min}_{\mathbf{n} \perp x} {\rm Var}[\mathbf{n}. \mathbf{\hat{S}}]$ and ${\langle \hat{S}^x \rangle^2 }$, neglecting the correlation between the two quantities. Likewise,  $(\hat{S}^x)^2$  and $(\hat{S}^y)^2$ are correlated in the evaluation of $\hat{M}_{xy}$; for simplicity, this correlation is ignored when estimating errors  for the dTWA averaging of a single disorder realization. 
Additional details on the error analysis for the scaling exponents are contained in the following section.

\begin{figure}
    \centering
    \includegraphics[width=\linewidth]{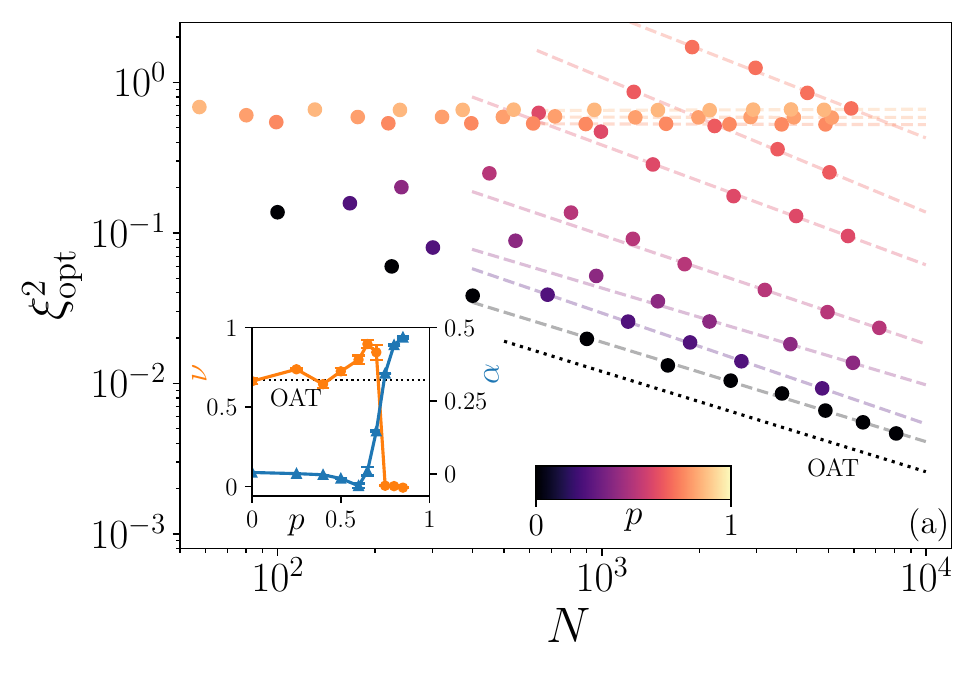}
    \caption{Optimal squeezing parameter $\xi_{\rm opt}^2$ vs $N$  for different vacancy probabilities $p$, where the colors range from $p=0$ to $p=0.85$ (legend) and we have set $\Delta = -1$.  The dotted line gives the scaling for OAT: $\xi_{\rm opt}^2\sim N^{-2/3}$. Error bars are included but are smaller than the marker size. Inset: $\nu$ (circles) vs $p$ and $\alpha$ vs $p$ (triangles), extracted from fits to the data in panels (a)(dashed lines). Error bars indicate uncertainty of the fit. Data is obtained from an average over 10 disorder realizations, with 6400 dTWA samples for each system.}
    \label{xi2_optdef}
\end{figure}

\section{Phase diagram data}
The data in Fig.~\ref{fig:phase_diagram}(a) and (b) are obtained for system sizes in the range $N \sim 1500-3000$, using 25 disorder realizations, with 1024 and 12800 dTWA samples per realization in panels (a) and (b), respectively. To demonstrate the scaling of $\xi^2_{\rm opt}$ in more detail, we additionally consider the cut $\Delta = -1$ of the phase diagram, for which we simulate substantially larger system sizes up to $N \sim 6000-8000$ with 10 disorder realizations. The corresponding results are shown in Fig.~\ref{fig:scaling}(a) of the main text. The larger system sizes reveal the impact of finite-size effects near the transition, shifting the estimated $p_c$ from 0.64(3) to 0.75(3). The former is plotted in Fig.~\ref{fig:phase_diagram}(a) and (b) for consistency with the other data points obtained for similar system sizes and disorder averaging. We therefore expect that the phase boundary inferred from $\xi_{\rm opt}^2$ would shift to higher values of $p_c$ if larger system sizes are considered, which for $\Delta < -1$ would reduce the discrepancy with the boundary obtained from the late-time magnetization.

As stated in the main text, in order to characterize the system size scaling of $\xi^2_{\rm opt} \sim N^{-\nu}$ we use a nuanced definition of $\xi^2_{\rm opt}$, in which we only consider minima for which the time $t_{\rm opt}$ diverges with the system size $t_{\rm opt} \sim N^{\mu}$, with $\mu > 0$. We also  exclude data for which $t_{\rm opt} < 5 J$, which further  assists to remove the non-scalable early time minima arising from local relaxation on time-scales of $\mathcal{O}(J)$, as is visible in the example of Fig.~\ref{fig:examples}(c). In the absence of minima fulfilling these requirements, as is typical in the disordered phase, we  take the global minimum. See Ref. \cite{Block_2024} for an extended discussion of the subtleties of extracting a clean scaling for $\xi^2_{\rm opt}$ and the related problem of estimating the phase boundary. 
 Fig.~\ref{fig:scaling}(a) of the main text shows results for the global minima, since it is visually more intuitive. In Fig.~\ref{xi2_optdef} we display results for the nuanced definition of $\xi^2_{\rm opt}$, which we use to extract the exponents. Note that there are points above $\xi^2 = 1$; these are local minima which trend towards becoming the true global minima at larger $N$, beyond what we can simulate directly. Nevertheless, the nuanced definition captures the scaling of these $p$-values, which are typically close to the transition.

Having established $\xi^2_{\rm opt}$ and its uncertainty (via the procedure in the previous section), the exponent $\nu$ can be estimated via the power-law fit (in log-log space). We use a weighted linear least-squares fit, which also provides uncertainties by assuming the likelihood is Gaussian near the optimal point. In the vicinity of the transition, a scalable minima may only emerge for the largest system sizes [as in Fig.~\ref{fig:examples}(c) of the main text]. If we do not observe at least three system sizes in this region, no exponent is estimated, since there is not enough data to establish a reasonable power-law fit. The exponent $\alpha$ is estimated from the late-time magnetization (see the subsequent section) in a similar way, albeit without the subtleties.

\begin{figure}[t]
\includegraphics[width=0.76\linewidth]{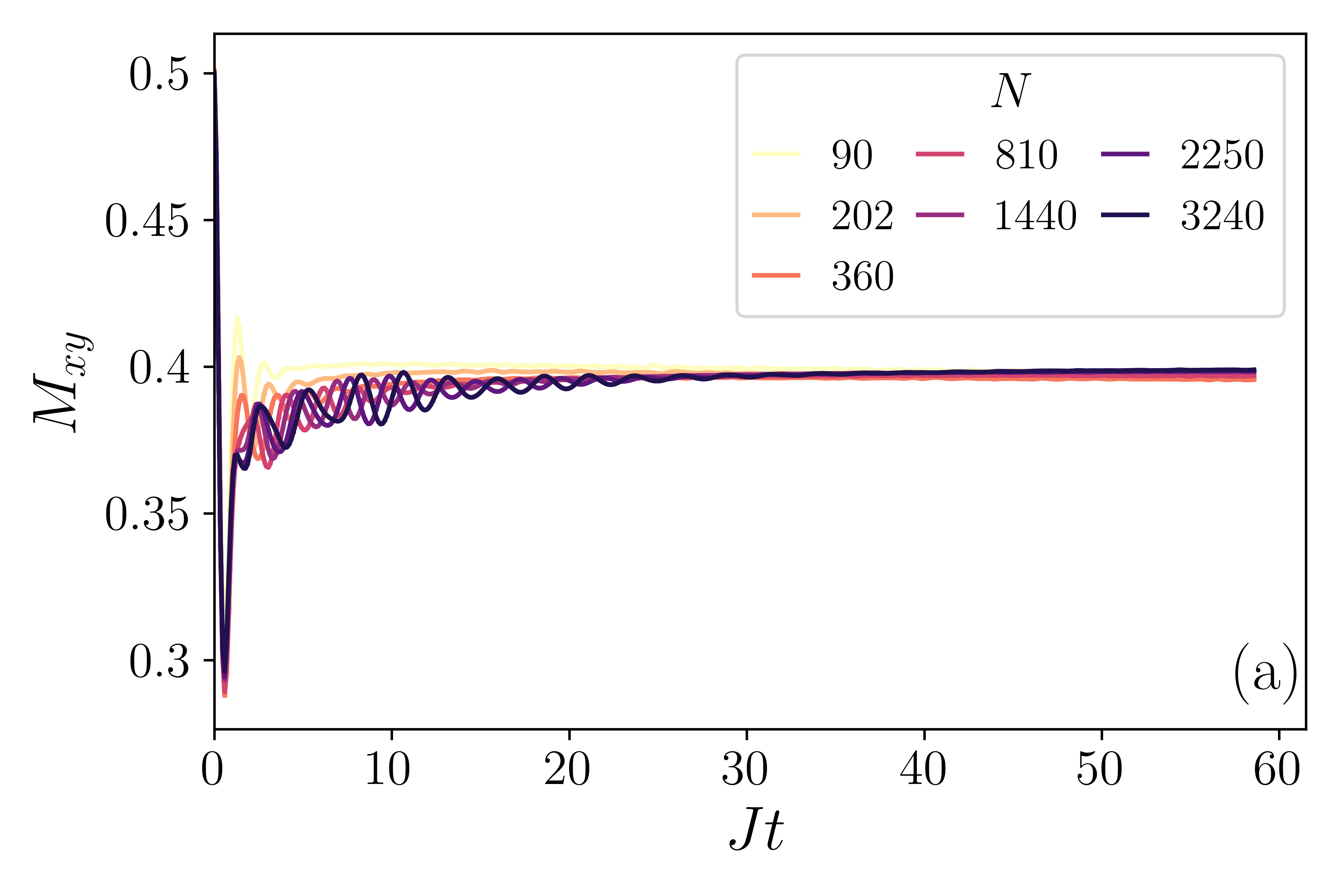}

    \includegraphics[width=0.76\linewidth]{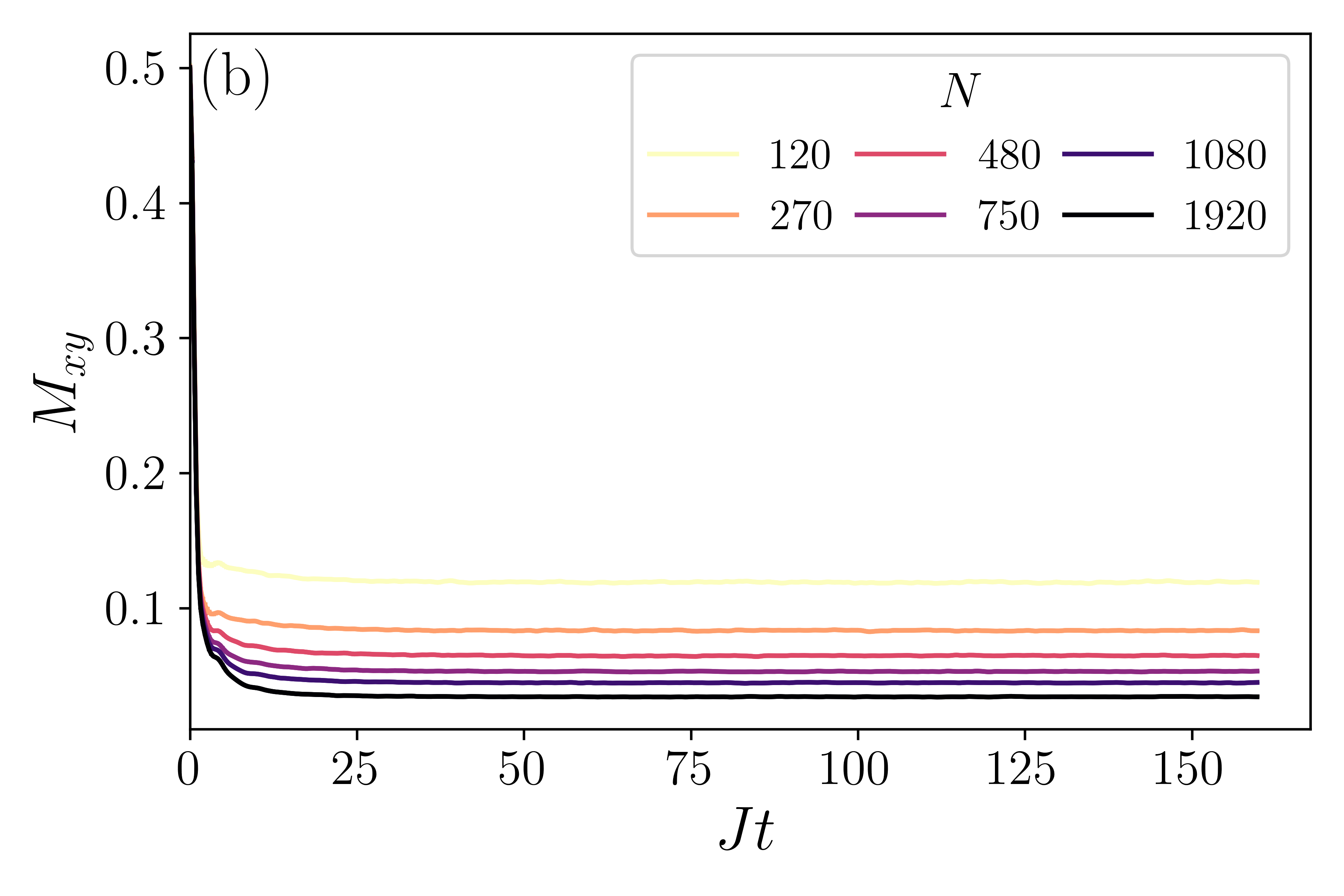}

    \includegraphics[width=0.76\linewidth]{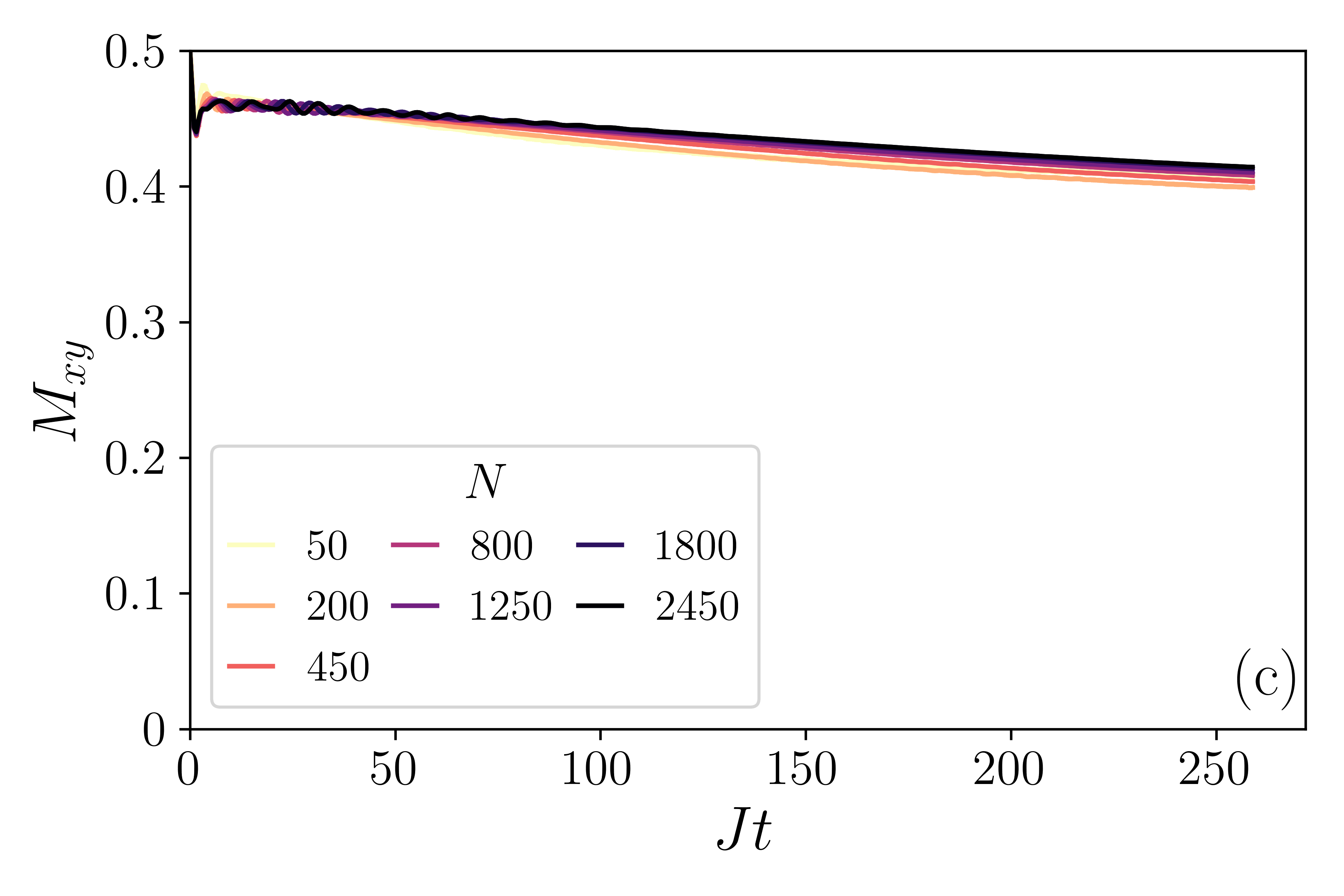}
    
    \includegraphics[width=0.76\linewidth]{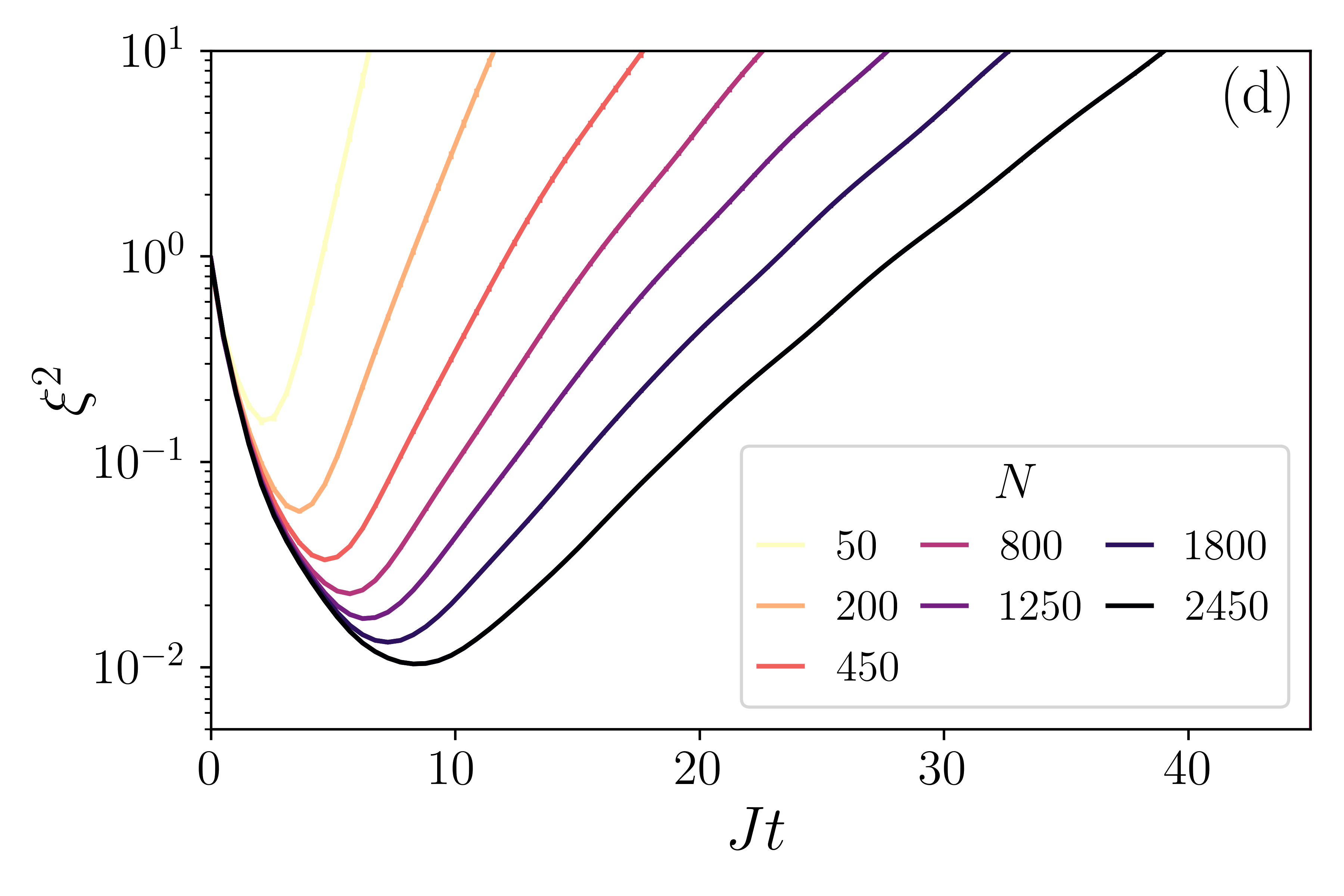}
    \caption{Magnetization $M_{xy}$ vs time for (a) $p = 0.1$ and (b) $p = 0.7$, for $\Delta = -2.0$. Data is shown for different system sizes; the disorder averaged system size $N = (1-p)L^2$ is shown in the legend. Data is obtained from an average over 25 disorder samples, with 1024 dTWA samples for each system. (c) Similar, but for $\Delta = 0$ with $p = 0.5$. The corresponding $\xi^2$ dynamics are shown in (d). Error bars are included in all plots but are typically smaller than the line width.}
    \label{fig:mag_long}
\end{figure}

 Fig.~\ref{fig:data_phase} shows the exponents (a) $\alpha$ vs $p$ and (b) $\nu$ vs $p$ for a range of $\Delta$ values. This data is used to construct Fig.~\ref{fig:phase_diagram} of the main text. The red dashed lines indicate the threshold values that we use to determine the critical $p$ values, $p_c$. Since we have data for a discrete set of $p$ values, linear interpolation between the data points is used (lines between points) to yield this value. The extracted $p_c$ are plotted as blue dots and red triangles in Fig.~\ref{fig:phase_diagram} of the main text. We also estimate the uncertainty in $p_c$ from both the vertical error bars (dTWA and disorder error)  and the discreteness of the  simulated $p$ grid in the x-direction. Since the discussion applies to both the magnetization  [Fig.~\ref{fig:phase_diagram}(a)] and  spin squeezing parameter [Fig.~\ref{fig:phase_diagram}(b)], we  use a generic symbol $y$ to denote either $\alpha$ or $\nu$ in what follows. In order to estimate the error on $p_c$ we first find the nearest points to the left ($p_l$) and right ($p_r$) of the threshold respectively, with corresponding $y$-values $y_{l/r}$, and standard errors $\delta y_{l/r}$. The uncertainty in $p_c$, which we denote $\delta p_c$ is derived from linear interpolation as 
\begin{align}
    (\delta p_c)^2 = \Big(\frac{p_r - p_l}{y_r - y_l}\Big)^2 \big\{ t^2 (\delta y_l)^2 + (1-t)^2 (\delta y_r)^2\big\} + \delta^2
\end{align}
where $t = \frac{p_c - p_l}{p_r - p_l}$. Even without uncertainty in the y-direction, the true crossing could lie anywhere within the bracketing interval, and we have conservatively included half the interval width $\delta= (p_r - p_l)/2$ as an additional uncertainty.

\section{Late-time magnetization}
To calculate the late-time magnetization displayed in Fig.~\ref{fig:scaling}(b) and Fig. \ref{fig:phase_diagram}(a) of the main text, in the ordered phase we evolve for at least $\mathcal{O}(10)$ times longer than is needed to reach the spin squeezing minimum ($t_{\rm opt}$) for the largest system sizes we consider. In the disordered phase we evolve for a comparably long time in units of $J$, although without reference to a time-scale associated with any spin squeezing minima.

 We extract the late-time magnetization $\overline{M}_{xy}$ from an average over the final $10\%$ of simulated times. Since the statistical errors across these times are correlated, we estimate the typical error from the average of the errors across the window.  In many of our simulations, the results correspond to time-scales of order $Jt \sim \mathcal{O}(10^2)-\mathcal{O}(10^3)$ (see Fig.~\ref{fig:mag_long} for an example).  
We typically take $\mathcal{O}(10)$ more samples when evaluating the spin squeezing parameter as compared to the magnetization; sample numbers are listed in the main text plots. This was chosen due to the smaller magnitude of $\xi^2$, as well as the shorter times of interest.

Fig.~\ref{fig:mag_long} shows the magnetization evaluated to late times for the case of $\Delta = -2.0$ and (a) $p = 0.1 $, (b) $p =0.7$. The former is in the ordered phase, while the latter is in the disordered phase, as is visible from the scaling of the late time values with the system size. For $-0.5 \leq \Delta \leq 0.5$ the magnetization can relax very slowly, and does not always fully converge to the steady state over accessible times. In this region we consider the magnetization at the longest available times. To determine whether the magnetization is relaxed we use two requirements. We divide the final $20\%$ of simulated times into two equally spaced regions of duration $T$. Defining the average magnetization of the earlier/later region as $m_{1/2}$,  we require that the difference between these is less than two times the typical standard error $\sigma$ for the final $20\%$ of times, i.e. $|m_1 - m_2|/\sigma < 2$. In addition, we take a linear fit to the data in the latter region (final $10\%$ of times) and project forward to see how much the magnetization will change if this gradient continued for an additional period of the same length. We require that this yields a change in the magnetization of less than 0.0025. In Fig.~\ref{fig:phase_diagram}(a) of the main text the non-converged results are indicated by  the hatched region.  An example of the slow relaxation is given in Fig.~\ref{fig:mag_long}(c). The corresponding $\xi^2$ dynamics are shown in Fig.~\ref{fig:mag_long}(d) and can be seen to reach a minimum over a much smaller time-scale. 

\begin{figure}
    \centering
    \includegraphics[width=\linewidth]{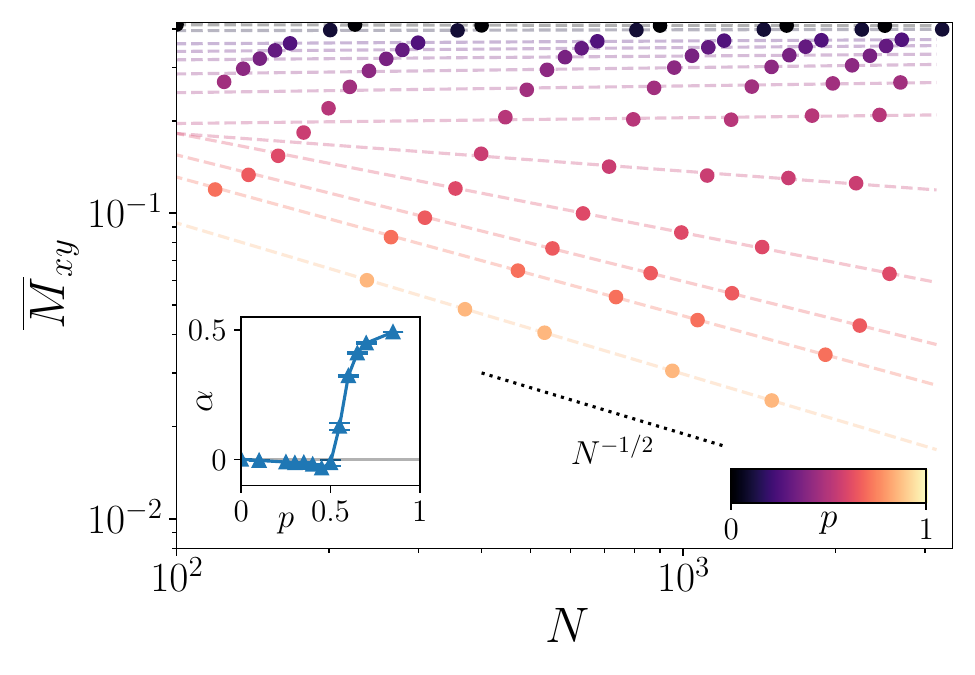}
 \caption{Late-time magnetization  $\overline{M}_{xy}$ vs system size $N$ for different $p$ values (legend), for $\Delta = -2.$ 
    Data is  obtained from an average over 25 disorder realizations, with 1024 dTWA samples for each system. Error bars are included but are smaller than the marker size. Inset: $\alpha$ vs $p$ , extracted from fits to the data in the main panel (dashed lines).
    Error bars indicate uncertainty of the fit.}
    \label{fig:mag_Delta_m2}
\end{figure}

Fig.~\ref{fig:mag_Delta_m2} shows $\overline{M}_{xy}$ vs $N$ in analogy to Fig.~\ref{fig:scaling}(b) in the main text, albeit for $\Delta = -2$. In this case there is more data in the disordered phase ($p_c \approx 0.5$), and the non-trivial power-law scaling is observed for many $p$ values. Larger $p$ values are seen to approach $\overline{M}_{xy} \sim N^{-1/2}$ scaling, as indicated from the exponents in the inset.

\section{Scaling of optimal squeezing time}
\begin{figure}
    \includegraphics[width = \linewidth]{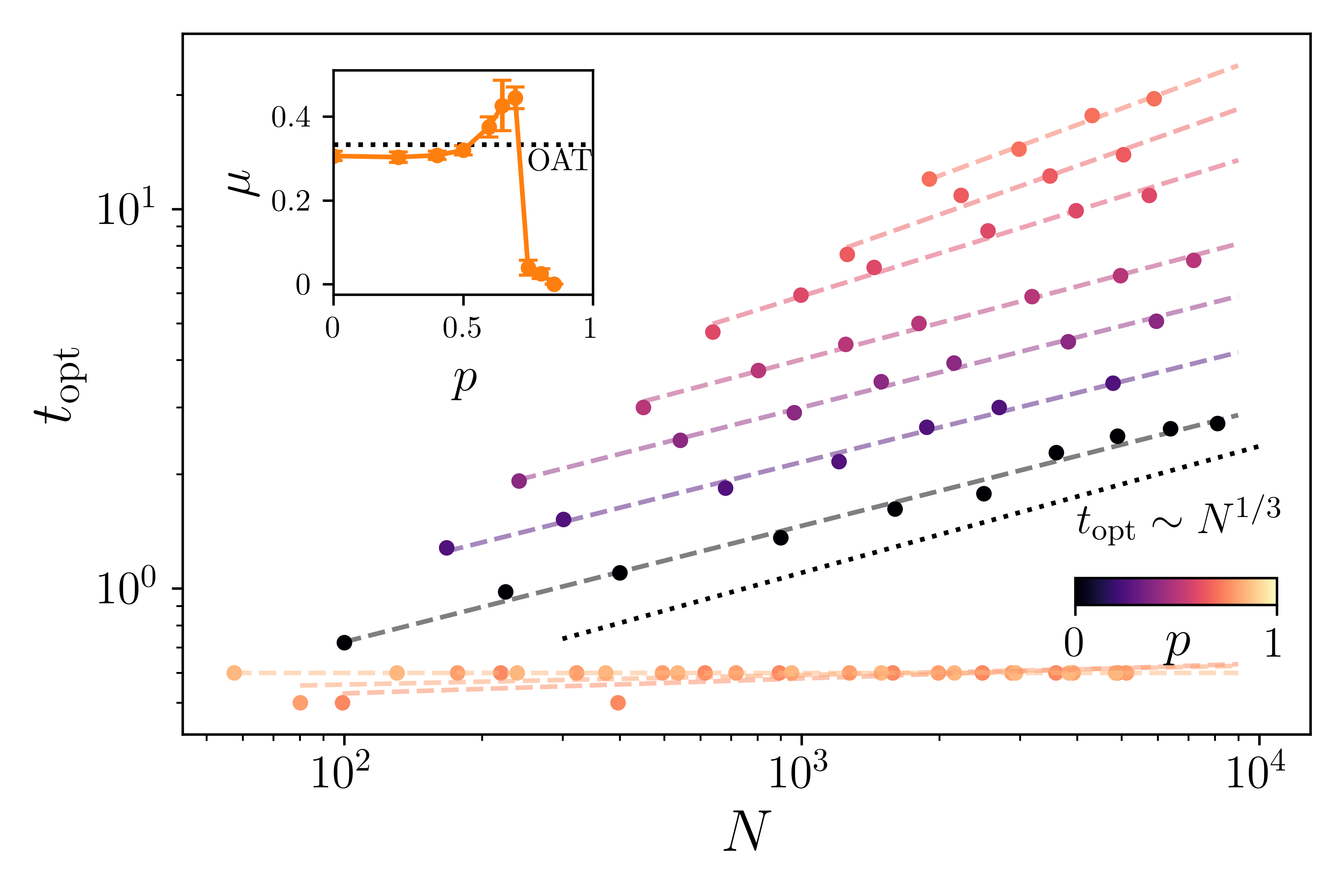}
    \caption{Optimal squeezing time $t_{\rm opt}$ vs system size $N$ for a range of $p$ values (legend), with $\Delta = -1$. Inset: exponent $\mu$ vs $p$ where $t_{\rm opt} \sim N^{\mu}$, extracted from a fit to the data in the main panel (dashed lines). }
    \label{fig:topt_vs_N}
\end{figure}

In Fig.~\ref{fig:topt_vs_N} we show the optimal squeezing time $t_{\rm opt}$ vs $N$ for a variety of $p$ values (legend), setting $\Delta = -1$. The data exhibit an approximate power-law $t_{\rm opt} \sim N^{\mu}.$ Up to the critical vacancy probability $p = p_c$ the scaling is comparable to the OAT result $t_{\rm opt} \sim N^{1/3},$ with larger deviations near the critical point expected to be finite-size effects.

\begin{figure}
    \includegraphics[width = \linewidth]{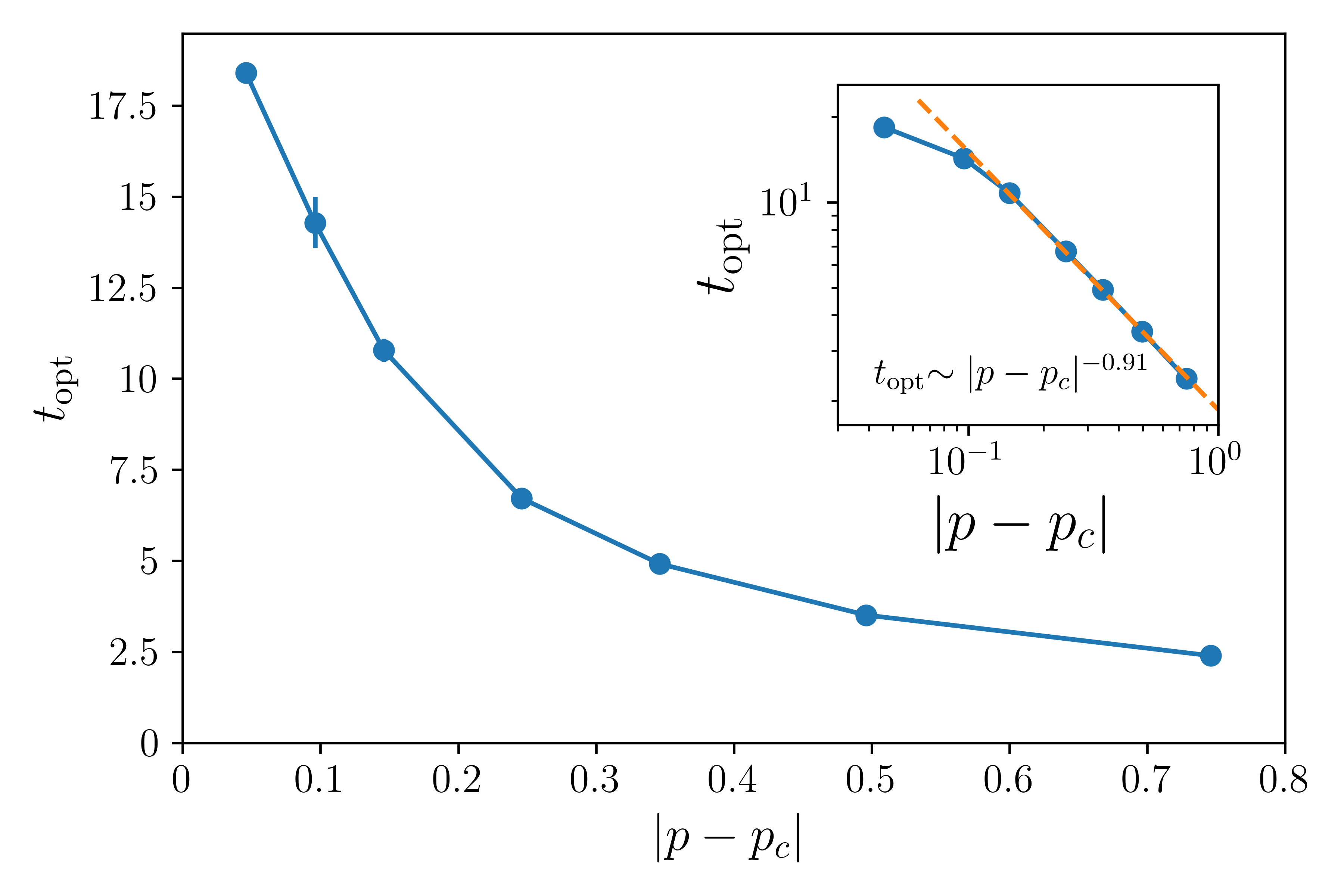}
    \caption{Optimal squeezing time $t_{\rm opt}$ vs $|p-p_c|$ for $N= 5 \times 10^3$ and $\Delta = -1$, estimated (with uncertainty) from the power-law fits in Fig.~\ref{fig:topt_vs_N}. Inset: same data in log-log space, demonstrating the scaling $t \sim |p-p_c|^{-\gamma}$ where $\gamma = 0.91(5)$ is estimated from the fit (dashed lines).}
    \label{fig:topt_vs_p}
\end{figure}

We now examine the scaling of $t_{\rm opt}$ as the phase transition is approached from the ordered phase. We use the power-law fits in Fig.~\ref{fig:topt_vs_N} to estimate the value of $t_{\rm opt}$ for each $p$ value at a constant $N$, shown in Fig.~\ref{fig:topt_vs_p}. For concreteness we have set $N =5 \times 10^3$. The data can be seen to exhibit a clear power-law scaling (inset) associated with critical slowing down: $t_{\rm opt} \sim |p-p_c|^{-\gamma}$, with $\gamma = 0.91(5)$. The uncertainty here indicates the range of values that are obtained by varying $N$ from $10^3$ to $10^4$ in this procedure, which exceeds the uncertainty on the fitting. The data point closest to the critical point $p_c = 0.75(3)$ is excluded when calculating the power-law fit. We expect that more accurate determination of the critical point would allow the power-law scaling to continue to smaller values of $|p-p_c|$. Ultimately, this requires simulating larger system sizes for the $p$ values closest to the critical point.  Combining the dependence on $p$ and $N$, we therefore observe the scaling $t_{\rm opt} \sim |p-p_c|^{-\gamma}N^{\mu}.$

\begin{figure}
    \centering
\includegraphics[width=\linewidth]{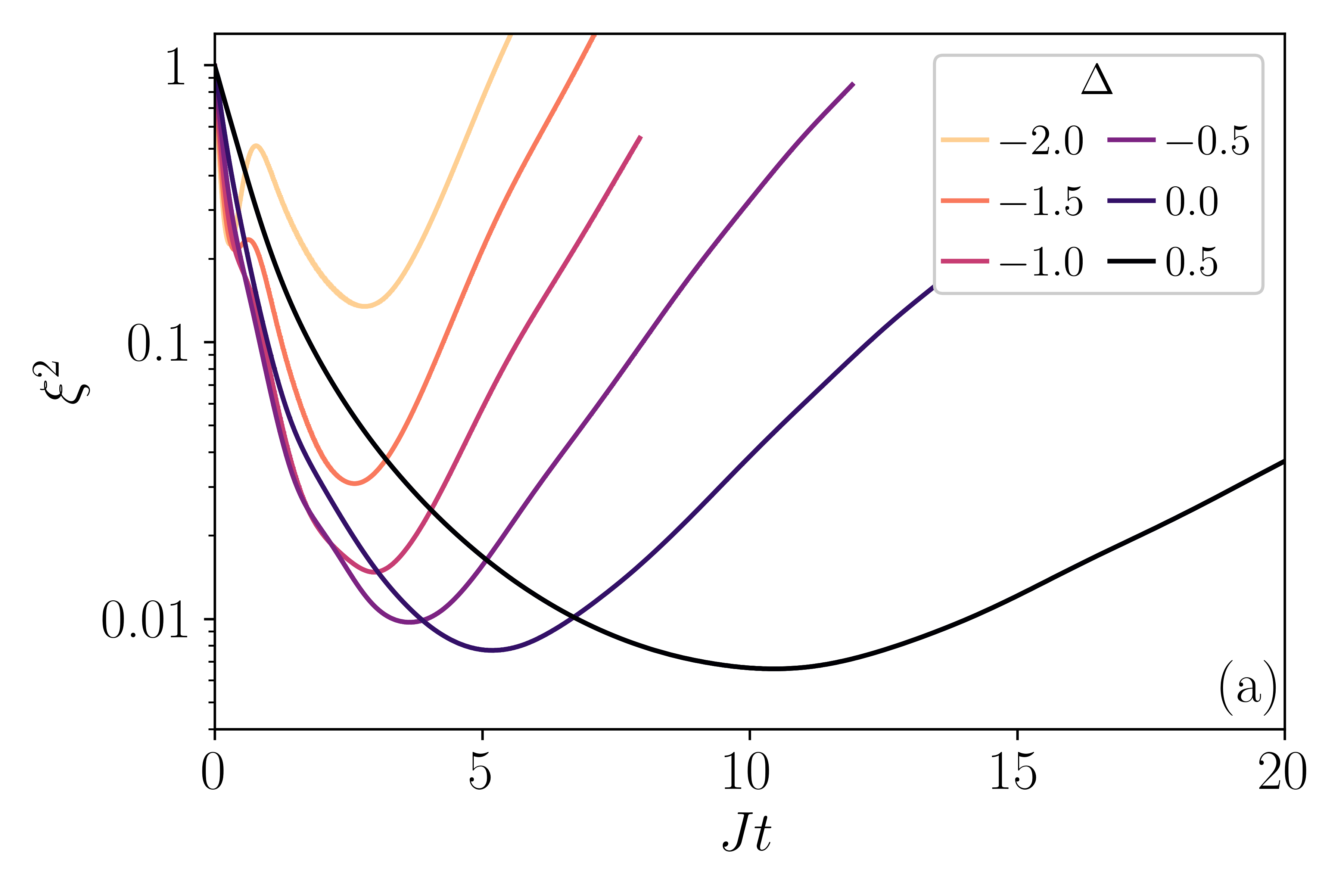}
\includegraphics[width=0.945\linewidth]{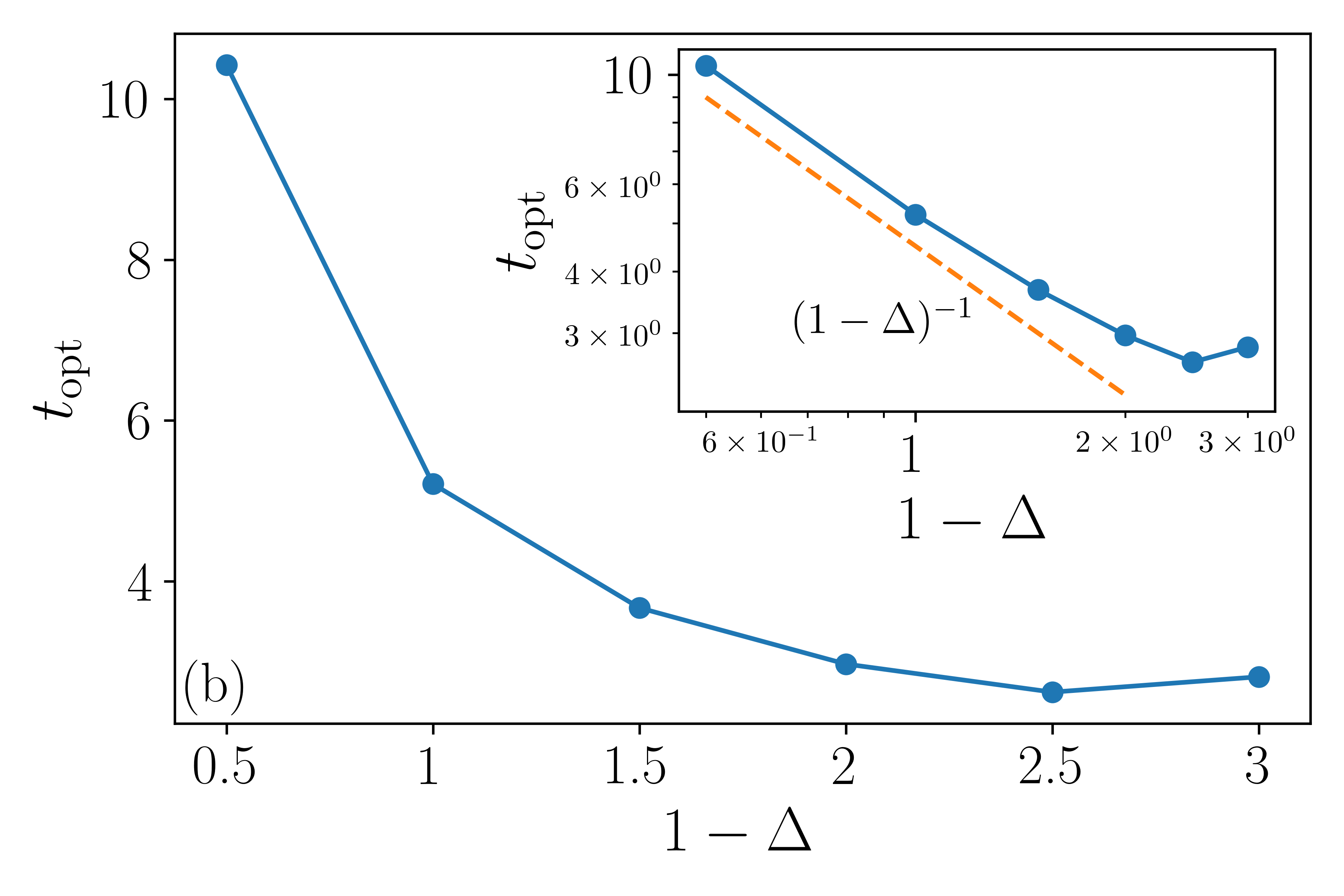}
    \caption{(a) Spin squeezing parameter $\xi^2$ vs time for varying $\Delta$ (legend), for constant $p = 0.25$ and $N = 1600$. Error bars are included but are typically smaller than the line width.  (b) Associated optimal squeezing time $t_{\rm opt}$ vs $1-\Delta$. Inset: data in log-log space. Dashed line indicates $t_{\rm opt} \sim (1-\Delta)^{-1}$. The results are obtained from 25 disorder realizations and 12800 dTWA samples.}
    \label{fig:slow}
\end{figure}

We end by briefly discussing the scaling of $t_{\rm opt}$ with $\Delta$. In Fig.~\ref{fig:slow}(a) we show the spin squeezing parameter $\xi^2$ vs time for a range of $\Delta$ values at fixed system size $N=1600$ and $p=0.25$. 
The minimum occurs at later times for the larger (more positive) $\Delta$. Fig.~\ref{fig:slow}(b) shows $t_{\rm opt}$ vs $1-\Delta$. The data is consistent with an approach to the rotor-spin-wave theory prediction for the disorder-free case of $t_{\rm opt} \sim (1-\Delta)^{-1}$ \cite{Roscilde2022,Roscilde2023}, suggesting a divergence of the squeezing time-scale at $\Delta=1$. While not explicitly relevant here (as $\alpha > D$ in our case), a similar scaling has been reported from perturbation theory in the genuinely long-range interacting case $\alpha < D$, where $D$ is the dimension \cite{Perlin2020}.

\section{ADDITIONAL DATA}
\begin{figure}
    \centering
\includegraphics[width=\linewidth]{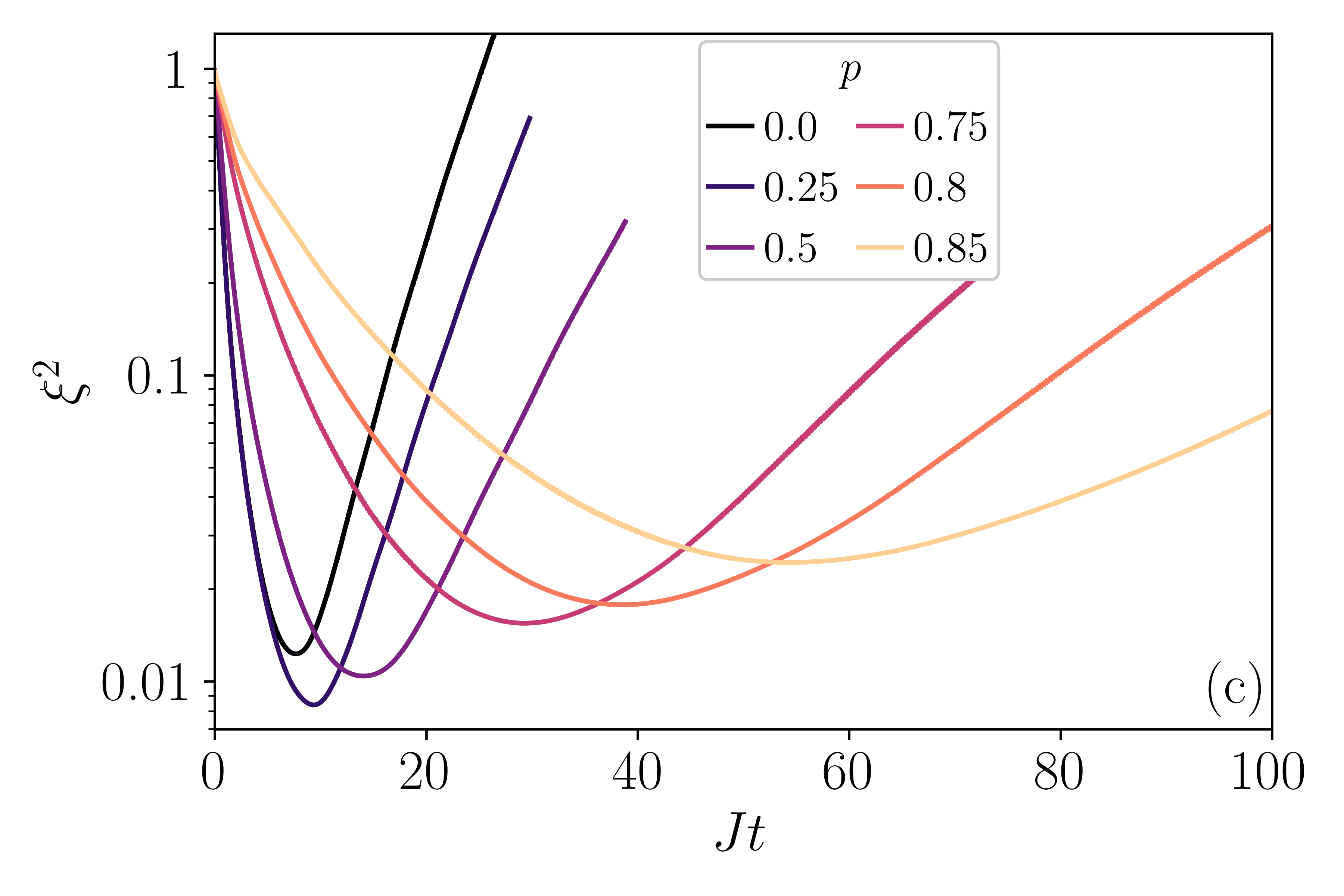}
   \caption{Spin squeezing parameter $\xi^2$ vs time  for varying $p$, setting $\Delta = 0.5$. The data corresponds to $N$ values in the range $N \in \{1450,1850 \}$, and is obtained from 25 disorder realizations and 12800 dTWA samples. Error bars are included but are typically smaller than the line width. }
        \label{fig:slow_p}
\end{figure}

In Fig.~\ref{fig:slow_p} we consider the spin squeezing parameter vs time for the case of $\Delta = 0.5$, this time varying $p$. The data for the different $p$ cases correspond to similar system sizes in the range $N \in \{1450,1850\}$. At larger $p$ values the time-scale to reach the minimum also diverges, which is associated with critical slowing down on approach to the finite temperature phase transition (discussed in the previous section). Based on the data in Fig. \ref{fig:phase_diagram}(b) of the main text, the critical value is in the region $p_c > 0.85$, which lies outside the parameters we simulate. 

\section{Distribution of effective interaction strengths}
\begin{figure*}
    \centering
    \includegraphics[width=\linewidth]{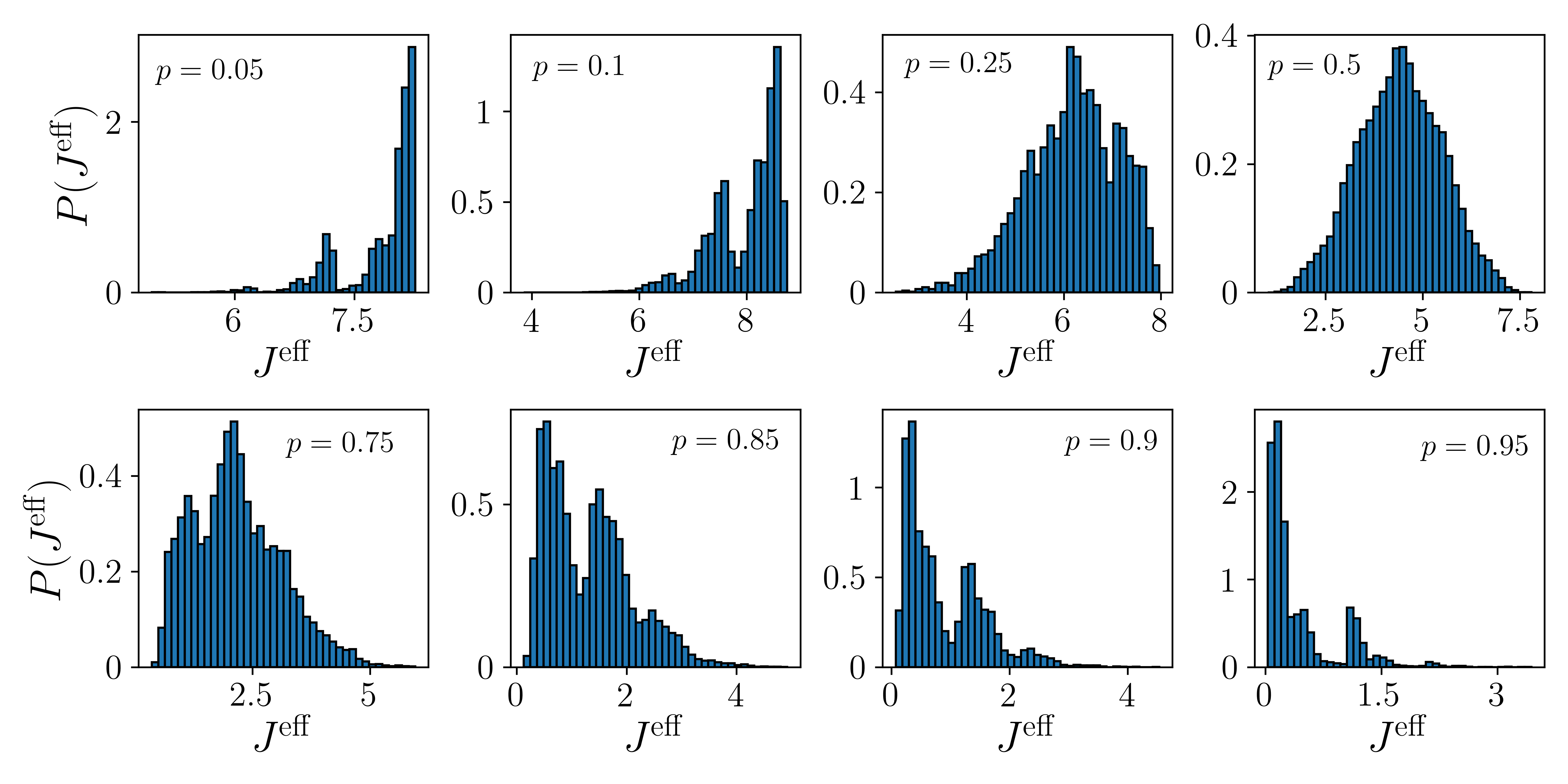}
    \caption{Distribution of effective interaction strengths $P(J^{\rm eff})$ for varying vacancy probability (labels). Data obtained from lattices with $N\sim \mathcal{O}(10^3)-\mathcal{O}(10^4)$ averaged over 25 disorder realizations.}
    \label{fig:distributions_large}
\end{figure*}

Fig.~\ref{fig:distributions_large} shows the effective interaction strength distribution $P(J^{\rm eff})$ for a range of $p$ values, in analogy to Fig.~\ref{fig:distributions} of the main text.

%
